\documentclass[aps,pra,showpacs,amsmath,amssymb,twocolumn,nofootinbib,superscriptaddress,10.5pt]{revtex4-2}

\usepackage[pdftex]{graphicx}

\usepackage{mathrsfs,amsmath,amssymb,amsthm,graphicx}
\usepackage{fancyhdr}

\usepackage[position=top,textfont=normalfont,singlelinecheck=false,justification=raggedright,format=hang,belowskip=0pt,font={small,it}]{subcaption}

\usepackage{braket}

\usepackage[colorlinks=true,
            linkcolor=blue,
            urlcolor=blue,
            citecolor=blue]{hyperref}
 
\begin{document}
\bibliographystyle{apsrev}

\title{Optimal Polarization-Entanglement Source: Frequency-Converted SPDC with Degeneracy, Indistinguishability, and Ultrahigh Purity That is Configurable Over a Large Spectral Range}
\author{Randy Lafler}
\affiliation{Air Force Research Laboratory, Directed Energy Directorate, Kirtland AFB, NM, United States}
\email[AFRL.RDSS.OrgMailbox@us.af.mil\\
Approved for public release; distribution is unlimited. \\
Public Affairs release approval AFRL-2021-4144]{}

\author{R. Nicholas Lanning}
\affiliation{Air Force Research Laboratory, Directed Energy Directorate, Kirtland AFB, NM, United States}




\date{\today}

\begin{abstract}
Modeling and simulations of entanglement-based quantum-networking protocols commonly assume perfect entangled states.
Some investigations have been performed, which show how imperfections cause the efficiency of the protocols to rapidly deteriorate.
For polarization-entangled states created by spontaneous-parametric down-conversion (SPDC), the fundamentals of phase matching lead to a trade-off problem for the optimal properties of the state.
We present a SPDC method, which circumvents the trade-off problem and allows one to obtain degeneracy, indistinguishability, and heralded-single-photon spectral purity greater than $99\%$ for any target SPDC wavelength in the visible and near-infrared spectrum. 
Therefore, our method can, in principle, generate polarization-entangled states that are optimal for polarization-entanglement-based quantum-networking protocols.
\end{abstract}

\maketitle 
\pagestyle{fancy}
\cfoot{Approved for public release; distribution is unlimited. Public Affairs release approval AFRL-2021-4144.}
\lhead{}
\chead{}
\rhead{\thepage}

\section{Introduction}
\label{sec: Introduction}
Quantum networking over long distances is an increasingly active area of research in the quantum optics community.
A driving force is the vision of a global-scale quantum internet that will provide fundamentally new internet technology by enabling quantum communication between any two points on Earth  \cite{wehner2018quantum,aspelmeyer2003long,boone2015entanglement,van2014quantum}. 
In this context, entanglement-based protocols such as quantum teleportation \cite{bennett1993teleporting} and entanglement swapping \cite{PhysRevLett.71.4287} are now considered quantum-networking primitives that may one day enable distributed quantum computing \cite{cuomo2020towards}, blind quantum computing \cite{barz2012demonstration}, and quantum-enhanced sensing \cite{gottesman2012longer,degen2017quantum} to name just a few applications. 
Practical implementations of any global-scale quantum-networking protocol will likely demand very high rates of entanglement consumption at distant network nodes.

Unless large quantities of entanglement can be established and stored for later use, very high rates of entanglement distribution will likely be required.
For global scales, the exponential loss in optical fiber prompts one to consider long-distance free-space links \cite{aspelmeyer2003long,boone2015entanglement}, which suffer only a quadratic loss due to geometric aperture-to-aperture coupling when the channel is described by the Friis equation \cite{friis1971introduction}.
In both cases, entanglement swapping and teleportation are used to distribute entanglement and information through the network, respectively. 
For polarization-based qubits \cite{kwiat1995new,kwiat1999ultrabright,shi2004generation,ursin2007entanglement, honjo2008entanglement,rangarajan2009optimizing}, the efficiency of distribution in the quantum network is hinged upon the efficiency of Bell-state measurements at intermediate nodes \cite{humble2007spectral,humble2008effects,jin2015highly}.
Therefore, ideally one uses perfect Bell states and even small deviations can quickly reduce the efficiency of distribution in a large network.
This is because the probability $\mathcal{P}$ of successfully establishing entanglement between the distant quantum nodes goes as $p^N$, where $p$ is the success probability of each entanglement swapping operation and $N$ is the number of quantum repeaters in the chain.  
Practically, this implies one needs $N$ identical entangled photon sources each having high polarization entanglement visibility, high heralded-single-photon purity, and indistinguishability in all degrees of freedom necessary for high interference visibility between photons from adjacent sources.

Quantum signal wavelength considerations further expand the networking trade space.
In the context of city-scale optical fiber-based networks, entanglement distribution can be conveniently established at telecom wavelengths, where optical fibers are less lossy and conventional polarization entanglement sources are intrinsically closer to ideal Bell states \cite{jin2015highly}.
In the context of free-space networking, the optimal wavelength for quantum communication is an enduring problem \cite{buttler2000daylight,gruneisen2015modeling,arteaga2019enabling,gruneisen2021adaptive,lanning2021quantum}.
For global-scale distances, the geometric aperture-to-aperture coupling can be approximated by the Friis equation and reveals a $1 / \lambda^2$ dependence \cite{friis1971introduction, alexander1997optical}.
Therefore, shorter wavelengths seem to have the advantage, especially in the case of space-space links.
On the other hand, space-to-Earth links include effects caused by the turbulent atmosphere and the Fried spatial-coherence equation reveals that the perceived turbulence is stronger at shorter wavelengths \cite{fried1966limiting}.
An analysis that carefully models these competing phenomenon has shown that certain shorter target wavelengths can provide a significant advantage over long wavelengths depending on atmospheric conditions and system parameters \cite{lanning2021quantum}.
The problem is that traditional sources of high-quality polarization-entangled qubits are in general difficult to develop and especially hard to achieve at arbitrary target wavelengths. 

\begin{figure*}[t]
\includegraphics[width=0.9\textwidth]{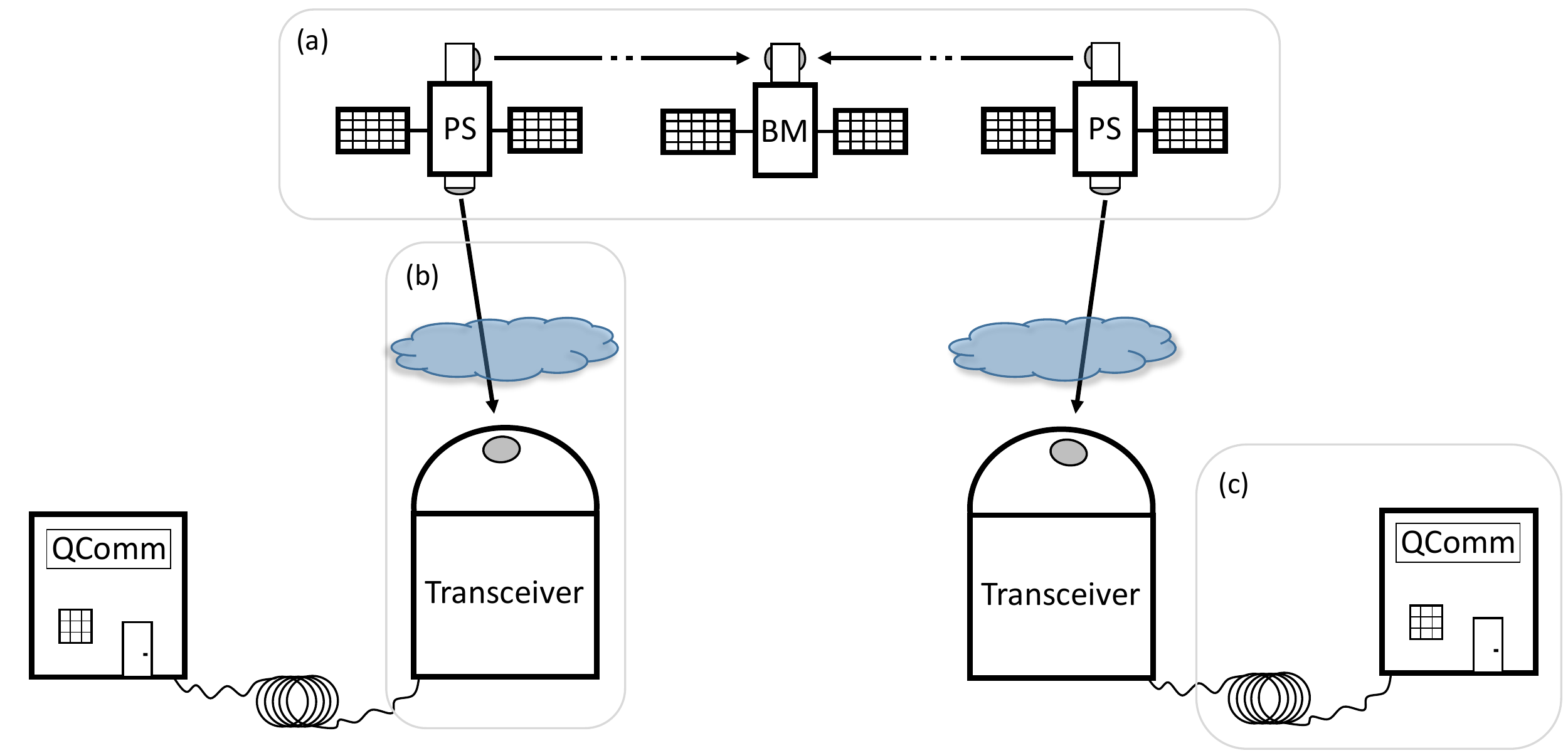}
\caption{
Example of a quantum-networking architecture where entanglement is distributed to two QComm terminals. Spaced-based photon sources (PSs) and Bell measurements (BMs) swap entanglement and transmit an entangled state to two ground transceivers via space-to-Earth downlinks. The ground transceiver may either directly transmit photons to the QComm terminals or perform wavelength transduction if necessary. Once entanglement is shared at the two terminals, qubits can be teleported from one terminal to the other for distributed and blind quantum computing, as an example. The schematic helps to illustrate the wavelength problem for global scale quantum networking, that is, each link has a different optimal wavelength which can be further complicated if quantum memories are to be used at intermediate and/or end nodes. 
}
\label{fig network_structures}
\end{figure*}  

The most common method for generating polarization-entangled photon pairs is spontaneous-parametric down-conversion (SPDC), where a pump photon spontaneously decays into a pair of lower-energy photons, commonly referred to as the signal and the idler \cite{kwiat1995new,kwiat1999ultrabright,grice1997spectral}. 
Typically, SPDC generates photon pairs that are strongly spatially and spectrally entangled, and consequently have low heralded-single-photon purity. 
The brute-force technique of tight spatial and spectral filtering can improve the heralded-single-photon spectral purity, but this comes at the expense of greatly reducing pair production \cite{mosley2008heralded,meyer2017limits}.
Therefore, it is unclear how this can be used to improve entanglement distribution rates.

A more graceful approach involves engineering the SPDC crystal to create photon pairs with ideal properties, thereby eliminating the need for spectral filtering.
However, there are trade-offs with this approach as well.
For example, to achieve high heralded-single-photon spectral purity one can engineer the group velocities of the pump, signal, and idler photons to obey the so-called group-velocity-matching (GVM) conditions, but this is possible only at certain wavelength combinations \cite{u2006generation}.
Therefore, high heralded-single-photon spectral purity often comes at the expense of forfeiting degeneracy and indistinguishability. 
Conversely, one may design a source to achieve degeneracy and indistinguishability, but this commonly comes at the expense of lower heralded-single-photon spectral purity.
One can  use aperiodic poling and domain engineering strategies to increase the heralded-single-photon spectral purity by removing certain spectral correlations, but only up to a limit when the GVM conditions are not satisfied \cite{u2006generation,dixon2013spectral,tambasco2016domain,graffitti2017pure,graffitti2018design}.
This trade-off problem can be detrimental to entanglement distribution protocols depending on the application and channel conditions. 

Consider a networking architecture including space-space, space-Earth, and terrestrial links; for example, an entanglement swapping link between two QComm terminals mediated by space-based sources and ground transceivers (see Fig.~\ref{fig network_structures}).
As stated previously, conventional polarization entangled sources near 1550 nm are intrinsically closer to ideal Bell states.
Therefore, network links based on fiber communication are immune to the trade-off problem [see Fig.~\ref{fig network_structures}(c)].
This along with the wavelength dependence of atmospheric transmission and sky radiance has encouraged some to conclude that 1550 nm is the optimal wavelength for free-space networking.
However, a rigorous analysis has shown that shorter wavelengths can significantly outperform longer wavelengths over atmospheric channels \cite{lanning2021quantum}.
There is an even stronger case for space-space links where there are no atmospheric effects and the $1 / \lambda^2$ dependence of the geometric aperture-to-aperture coupling is the dominant phenomenon [see Fig.~\ref{fig network_structures}(a)].   
Therefore, we question whether degenerate operation at 1550 nm is a desirable choice.
One can also consider nondegenerate sources in free-space entanglement-swapping links.
For example, one can perform swapping through a satellite constellation by alternating the wavelengths of the Bell measurements in the chain, but this introduces alternating link efficiencies. 
Consider the convenient wavelength combination in potassium titanyl phosphate (KTP), which intrinsically yields high heralded-single-photon spectral purity for down-conversion to 532 and 1550 nm.
Examining the architecture in Ref.~\cite{khatri2021spooky}, one will find that even in the best-case scenario the 1550-nm link would be approximately 10 db less efficient than the 532-nm link due to geometric aperture-to-aperture coupling loss using 10-cm apertures. 
This inefficiency is compounded as the number of necessary links grows.
For space-Earth links [see Fig.~\ref{fig network_structures}(b)], the entangled state will traverse the Earth's atmosphere and it therefore seems natural to choose wavelengths degenerate at an optimal wavelength \cite{lanning2021quantum}.
Altogether, this suggests that degeneracy at a shorter wavelength should be considered for the space-space and space-to-Earth links in Figs.~\ref{fig network_structures}(a) and \ref{fig network_structures}(b).
This would require wavelength transduction at the ground transceiver to convert to 1550 nm for telecom transmission, but this may already be the case in order to utilize quantum memory.
Thus, we see that the wavelength and heralded-single-photon spectral purity trade-off issue is a very relevant problem, which is directly impeding the development of high rate quantum networks.

We present a method using SPDC and sum-frequency conversion in tandem to create degenerate, indistinguishable, and polarization-entangled photon pairs over a large range of target down-conversion wavelengths.
Furthermore, we show our scheme in principal can achieve ultrahigh heralded-single-photon spectral purity ($>$ 99$\%$) over much of the visible and near-infrared (NIR) spectrum by selecting the optimal pump and crystal phase-matching bandwidths and the optimal crystal phase-matching configuration.
This is accomplished with broad spectral filtering and even without spectral filtering for the case of aperiodic poling and domain engineering.
Thus, our frequency-converted SPDC scheme provides the opportunity to realize the ideal entangled photon source, which is so commonly assumed in the theoretical descriptions of quantum protocols, e.g., entanglement swapping and teleportation.
For more details, see Refs.~\cite{humble2007spectral} and \cite{humble2008effects} for a description of the effect of spectral purity on quantum teleportation and entanglement swapping.

The paper is organized as follows.  
In Sec.~\ref{sec: Constraints SPDC} we briefly review SPDC theory, emphasizing the constraints on the heralded-single-photon spectral purity.  
In Sec.~\ref{sec: FC SPDC} we introduce our frequency-converted SPDC (FC-SPDC) scheme, and we discuss how its properties can be chosen to achieve ultrahigh heralded-single-photon spectral purity. 
In Sec.~\ref{sec: entangled fc SPDC} we present a scheme for creating optimal polarization entanglement.  
In Sec.~\ref{sec: Results} we present an optimization process using the nonlinear crystals KTP, and lithium niobate (LN), and investigate performance over a wide range of target down-conversion wavelengths.  
Thus, we show how one can configure our scheme to obtain degeneracy, indistinguishability, and heralded-single-photon spectral purity greater than $99\%$ for any target wavelength in the visible and the NIR.

\section{Purity Constraints Of Traditional Biphoton Sources}
\label{sec: Constraints SPDC}
In this section we briefly review the theory of SPDC and discuss the constraints on the heralded-single-photon spectral purity \cite{keller1997theory,grice1997spectral,yang2008spontaneous,u2006generation}.
In SPDC, photons from a high-powered pump beam spontaneously decay into pairs of photons, which often are strongly spectrally entangled.  
This spectral entanglement means heralding with one of the pair projects the other photon into a mixed-spectral state.
This is a problem for protocols involving the interference of photons from independent sources because the mixedness of the states degrades the interference visibility.
We now briefly show how one can calculate the state and investigate its heralded-single-photon spectral purity. 

The SPDC state $\ket{\Psi_{\mathrm{SPDC}}}$ can be written as the unitary evolution of the operator $U_{\mathrm{SPDC}}$ on the vacuum state $\ket{0}$
\begin{equation}
\label{Psi SPDC}
\ket{\Psi_{\mathrm{SPDC}}} = U_{\mathrm{SPDC}}\ket{0}.
\end{equation}
We expand $U_{\mathrm{SPDC}}$ to first order and ignore the vacuum contribution, thus keeping only the so-called biphoton term
\begin{equation}
\label{U_SPDC}
U_{\mathrm{SPDC}} = \int\int f_{\mathrm{JSA}}(\omega_s,\omega_i)\hat{a}_j^\dagger(\omega_{s})\hat{a}_k^\dagger(\omega_{i}) d\omega_sd\omega_i,
\end{equation}
where $\hat{a}_j^\dagger(\omega_s)$ and $\hat{a}_k^\dagger(\omega_i)$ are the creation operators for the signal and the idler photons at the frequencies $\omega_s$ and $\omega_i$ and with the polarizations $j$ and $k$, respectively, and $f_{\mathrm{JSA}}$ is the joint spectral amplitude (JSA)
\begin{equation}
\label{equ: JSA}
f_{\mathrm{JSA}}(\omega_s,\omega_i) = \alpha(\omega_s+\omega_i)\Phi(\omega_s,\omega_i),
\end{equation}
where $\alpha$ is the pump-envelope function (PEF) and $\Phi$ is the phase-matching function (PMF).
The PEF contains the spectral distribution of the pump and enforces energy conservation, that is, $\omega_p=\omega_s+\omega_i$, where $\omega_p$ is the frequency of the pump photon.
For a Fourier-transform-limited pulse the PEF can be modeled by an antidiagonal Gaussian, where the FWHM spectral width $\Delta\omega$ is equal to $0.44/\Delta t$ and $\Delta t$ is the FWHM pulse duration [for an example see Fig.~\ref{fig Visual Lambda Space}(a)].
The PMF is given by 
\begin{equation}
\label{equ: PMF}
\Phi(\omega,\omega^\prime) = \int_{-L/2}^{L/2} g(z)e^{-\Delta k(\omega,\omega^\prime)z}dz, 
\end{equation}
where $g(z)$ is the poling structure function, $\Delta k$ is the phase mismatch, and $L$ is the length of the nonlinear crystal.  
The poling structure function $g(z)$ can be $\pm 1$ depending on the orientation of the ferroelectric domain at $z$.  
When $g(z)$ alternates periodically, the resulting PMF is a sinc function.
The sidebands of the sinc function can induce spectral correlations in $\ket{\Psi_{\mathrm{SPDC}}}$.
These spectral correlations typically persist even for optimal GVM conditions, thus limiting the heralded-single-photon spectral purity.
Alternately, using domain-engineering strategies, $g(z)$ can be engineered to remove the sidebands and generate a Gaussian PMF \cite{u2006generation,tambasco2016domain,graffitti2017pure,graffitti2018design} [see Fig.~\ref{fig Visual Lambda Space}(c) for an example of a Gaussian PMF].
The phase mismatch $\Delta k$ encodes the crystal properties and enforces momentum conservation according to
\begin{equation}
\label{equ: phase-mismatch}
\Delta k=k_p-k_s-k_i-\frac{2\pi m}{\Lambda},
\end{equation}
where $k_j=n_j(\omega_j) \, \omega_j/c$ is the wave number for the $j=(p,s,i)$ wave, $\Lambda$ is the poling period for quasi-phase-matching, and $m$ is the phase-matching order.  

\begin{figure}[t!]
\captionsetup{width=\linewidth}
\includegraphics[width=\linewidth]{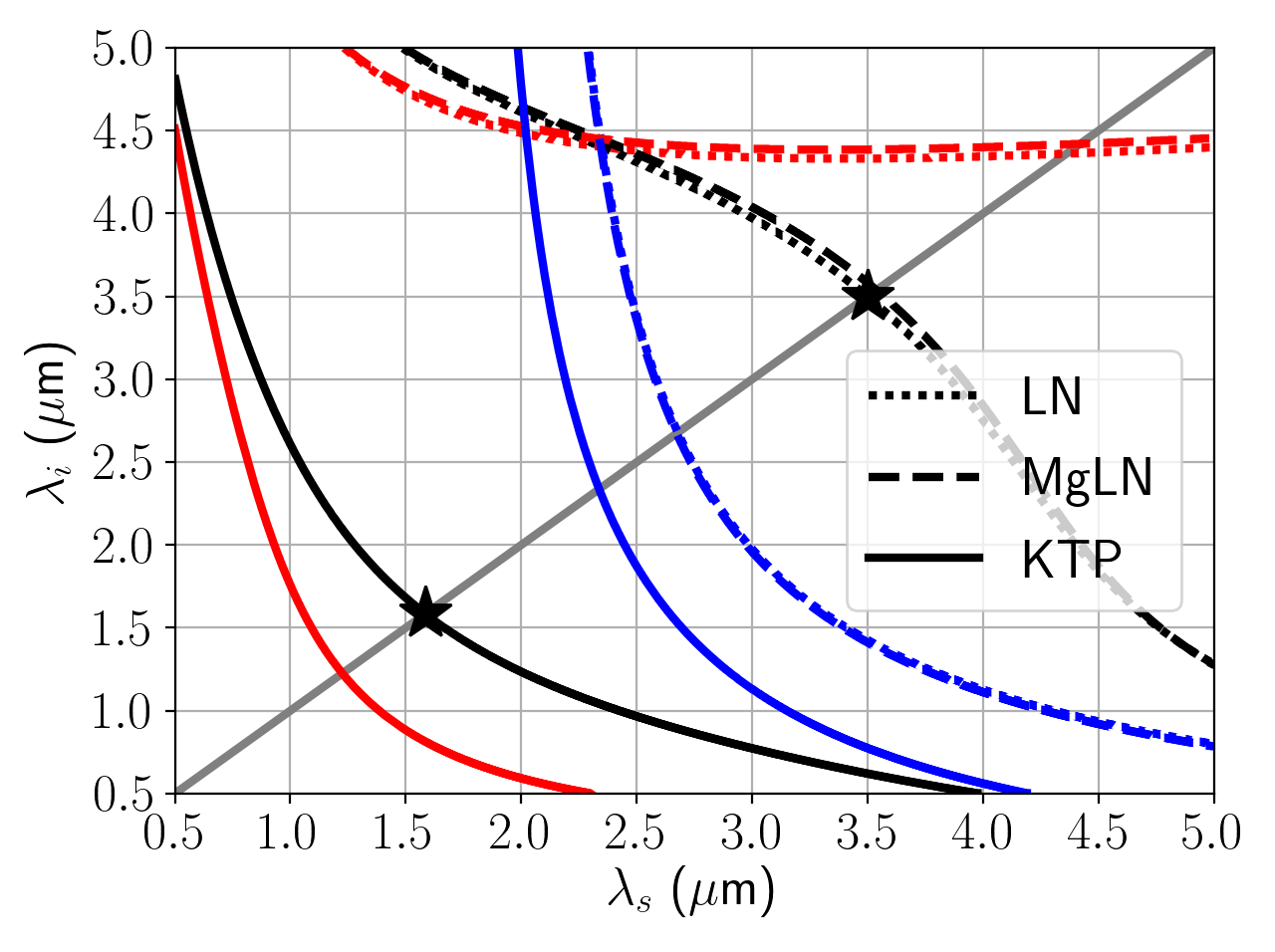}
\caption{
Group-velocity-matching conditions for type 2 phase matching in KTP, LN, and MgLN, which reveal the trade-off problems we resolve in this work.
The red, black, and blue curves trace out unit spectral purity states with vertically, circularly, and horizontally oriented joint spectral amplitudes, respectively.  
Intersection with the gray line across the diagonal indicates degeneracy conditions for the signal and idler wavelengths $\lambda_s$ and $\lambda_i$, respectively.  
One should notice the only states that are simultaneously degenerate, indistinguishable, and have high purity are at approximately 1550 nm, and approximately 3500 nm, indicated by the black stars.
}
\label{fig GVC}
\end{figure} %

\begin{figure}[t]
\captionsetup{width=\linewidth}
\includegraphics[width=\linewidth]{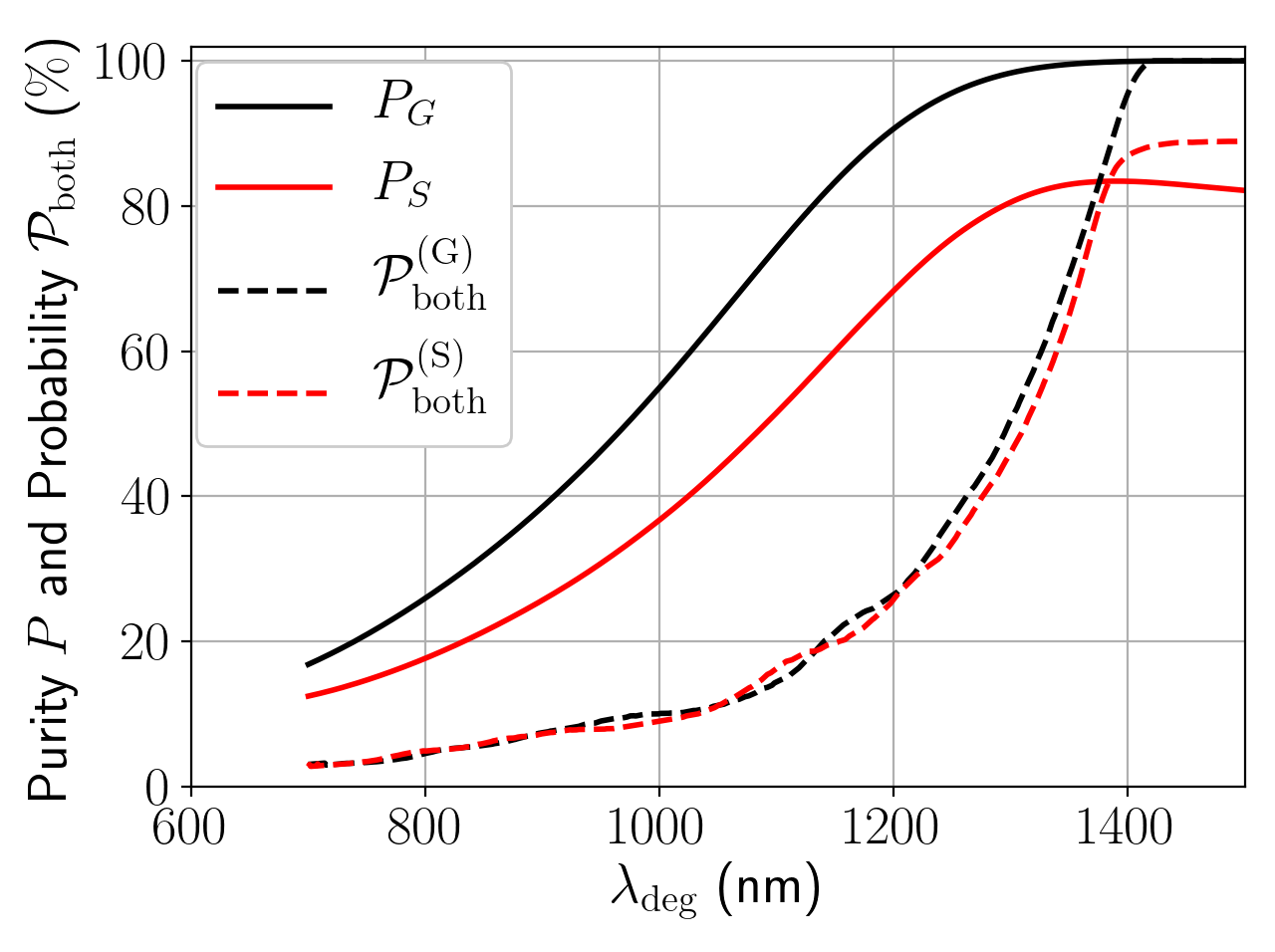}
\caption{
Degenerate PPKTP when using Gaussian (black) and sinc (red) phase matching, the heralded-single-photon spectral purity (solid, $\mathrm{P_\mathrm{{(i)}}}$), and the probability that both photons of a pair make it through identical top-hat filters that are just narrow enough to increase the purity to $99\%$ (dashed, $\mathcal{P}_{\mathrm{both}}^{(i)}$), where G and S refer to Gaussian and sinc phase matching, respectively.
}
\label{fig Deg KTP Filter}
\end{figure} 

A convenient measure of the spectral entanglement comes by means of the Schmidt decomposition of $f_{\mathrm{JSA}}(\omega_s,\omega_i)$ \cite{graffitti2018design,zielnicki2018joint}.
When the JSA is discretized over $\omega_s$ and $\omega_i$ one can write a matrix component as
\begin{equation}
\label{equ: Schmidt Decomp}
\big[f_{\mathrm{JSA}}\big]_{qr} = \sum_m\lambda_mu_{qm}v_{mr},
\end{equation}
where the Schmidt coefficients $\lambda_m$ satisfy the normalization condition $\sum\lambda_m^2=1$, and $u_{ms}$ and $v_{mi}$ are the Schmidt vectors.  
The Schmidt number $K=\sum_m 1/ \lambda_m^4$ is an indicator of spectral entanglement, that is, $\ket{\Psi_{\mathrm{SPDC}}}$ exhibits a high degree of spectral entanglement when $K>>1$.
Conveniently, the heralded-single-photon spectral purity, which we will refer to simply as the purity $P$, is \cite{u2006generation}
\begin{equation}
\label{equ: Purity}
P = 1/K,
\end{equation}
and thus one can see how spectral entanglement degrades purity.
The purity $P$ is also related to the frequency-space orientation of the JSA, that is, the purity $P$ is smallest when the function $f_{\mathrm{JSA}}$ is oriented along the diagonal or antidiagonal.  
Conversely, when $K=1$, spectral entanglement vanishes, the purity $P$ is unity, and the JSA is oriented horizontally, vertically, or circularly with no sidebands.  
When the bandwidth of the PEF $\alpha$ is greater or equal to the bandwidth of the PMF $\Phi$, the orientation of the JSA is dominated by the orientation of $\Phi$, which in turn is determined by the orientation of $\Delta k(\omega_s,\omega_i)$.  
Approximating Eq. (\ref{equ: phase-mismatch}) to first order in $\omega$ and setting $\Delta k=0$, one finds that the JSA is oriented along the line
\begin{equation}
\label{equ: dk exp}
\Delta\omega_i=\frac{k_p^\prime-k_s^\prime}{k_i^\prime-k_p^\prime}\Delta\omega_s,
\end{equation}
where 
\begin{equation}
\label{equ: dk exp2}
\begin{split}
\Delta\omega_j &= \omega_j-\bar{\omega}_j,\\
k_j^\prime &= \frac{dk}{d\omega}\Big|_{\omega_j=\bar{\omega}_j},
\end{split}
\end{equation}
and $\bar{\omega}_j$ is the central frequency of the $j=(p,s,i)$ photon.  
The group velocity $V_g$ of the photons are equal to $V_j=1/k_j^\prime$.  
From Eqs. (\ref{equ: dk exp}) and (\ref{equ: dk exp2}) one can find that the GVM conditions for a vertical, horizontal, or circular JSA are $k_p^\prime=k_i^\prime$, $k_p^\prime=k_s^\prime$, or $k_p^\prime=(k_i^\prime+k_s^\prime)/2$, respectively~\cite{u2006generation,kaneda2016heralded}.
These are sometimes further categorized as asymmetric GVM (AGVM) for the horizontal and the vertical conditions \cite{kaneda2016heralded}, and as symmetric GVM (SGVM) for the circular condition.  
One should notice for the AGVM conditions the $\alpha$ bandwidth must be larger than the $\Phi$ bandwidth, otherwise the JSA is oriented along $\alpha$ (antidiagonal) instead of along $\Phi$ (horizontal or vertical). 
Similarly, the bandwidths of $\alpha$ and $\Phi$ must be precisely matched for the SGVM condition. 

Another relevant quantity is the spectral indistinguishability $I$, which we calculate from $f_{\mathrm{JSA}}$:
\begin{equation}
\label{equ Indist}
I = \frac{\int \int f_{\mathrm{JSA}}(\omega_s,\omega_i)f^\dagger_{\mathrm{JSA}}(\omega_s,\omega_i)d\omega_sd\omega_i}{\int \int (f_{\mathrm{JSA}}(\omega_s,\omega_i))^2d\omega_sd\omega_i},
\end{equation}
where $f_{\mathrm{JSA}}^\dagger$ is the conjugate transpose of $f_{\mathrm{JSA}}$.

Figure \ref{fig GVC} shows the GVM conditions in wavelength space for type 2 SPDC phase matching in KTP, and LN.  
By type 2 SPDC phase matching we are referring to the pump, signal, and idler photons polarized along the fast, fast, and slow axis of the nonlinear crystal, respectively.
Assuming Gaussian PMFs, the red, black, and blue curves trace out unit-purity states with vertically, circularly, and horizontally oriented $f_{\mathrm{JSA}}$, respectively.  
The markers indicate states that are degenerate.
Since the states along the red and the blue curves have AGVM conditions, their $f_{\mathrm{JSA}}$ are elliptical and the signal and idler photons are distinguishable, even for the degenerate states.
In fact, the only states simultaneously possessing high purity, indistinguishability, and degeneracy are the states at approximately1550 nm, and approximately 3500 nm, indicated by the black markers.

In other nonlinear crystals the situation is similar, namely, states with high purity, indistinguishability, and degeneracy occur only at a single wavelength, typically in the IR \cite{laudenbach2016modelling}.
In Refs. \cite{jin2016spectrally,jin2019theoretical,jin2020spectrally} the authors locate these states in the isomorphs of KTP, potassium dideuterium phosphate, and beta barium borate, adding to the list of wavelengths at which these states can be obtained. 
However, there is still a broad range of degeneracy wavelengths that cannot be achieved, and thus there is a trade-off; one must sacrifice high purity for the sake of degeneracy and indistinguishability, or one must sacrifice degeneracy and indistinguishability for the sake of high purity.
Making matters worse, when phase-matching in bulk or quasi-phase-matching in periodically poled (PP) nonlinear crystals, the sinc sidebands of $\Phi$ limit the purity to the low $80$s and $90$s for SGVM and AGVM conditions, respectively.
To illustrate this we plot the purity for degenerate PPKTP for both Gaussian and sinc phase matching. 
In Fig.~\ref{fig Deg KTP Filter} one can see that the sinc (solid red) curve is limited to just above 80$\%$ whereas the Gaussian (solid black) curve approaches unit purity at long degeneracy wavelengths $\lambda_{\mathrm{deg}}$ \cite{u2006generation,jin2013widely}.

The brute-force method to overcome these trade-offs is to use tight spectral filtering, but this comes at the expense of greatly reducing the brightness of the source and the heralding efficiency \cite{mosley2008heralded,meyer2017limits}.  
In Fig.~\ref{fig Deg KTP Filter}, the black and red dashed curves are the probability that both down-conversion photons pass through top-hat filters that are narrow enough to increase the purity to $99\%$: 
\begin{equation}\label{eq:P_BOTH}
\mathcal{P}_{\mathrm{both}}^{(i)} = | f_{\mathrm{filt}}| ^2/| f_{\mathrm{JSA}}| ^2,
\end{equation}
where $f_{\mathrm{filt}}$ is the filtered JSA for conventional SPDC and $i$ indicates either Gaussian or sinc phase matching, respectively.
One can see that at short wavelengths very narrow filtering is required to obtain high purity.
In fact, below 800 nm only a few percent of the total down-converted photon flux makes it through the filters.
This is the trade-off when using tight spectral filtering, which may limit the rate of entanglement distribution of a quantum network.

In the next section we present how our FC-SPDC scheme can circumvent the trade-offs discussed in this section.
In effect, it is a more elegant spectral filtering strategy that relies on the overlap of the phase-matching functions and pump envelope functions involved in the interaction.
The conversion efficiency of this process is discussed in comparison to the brute-force spectral-filtering method, particularly in the context of $\mathcal{P}_{\mathrm{both}}^{(i)}$ and maintaining heralding efficiency. 
Furthermore, efficiencies related to transverse spatial mode overlap are also discussed.

\section{Frequency-converted SPDC}
\label{sec: FC SPDC}
\begin{figure}[t!]
\captionsetup{width=\linewidth}	
\includegraphics[width=\linewidth]{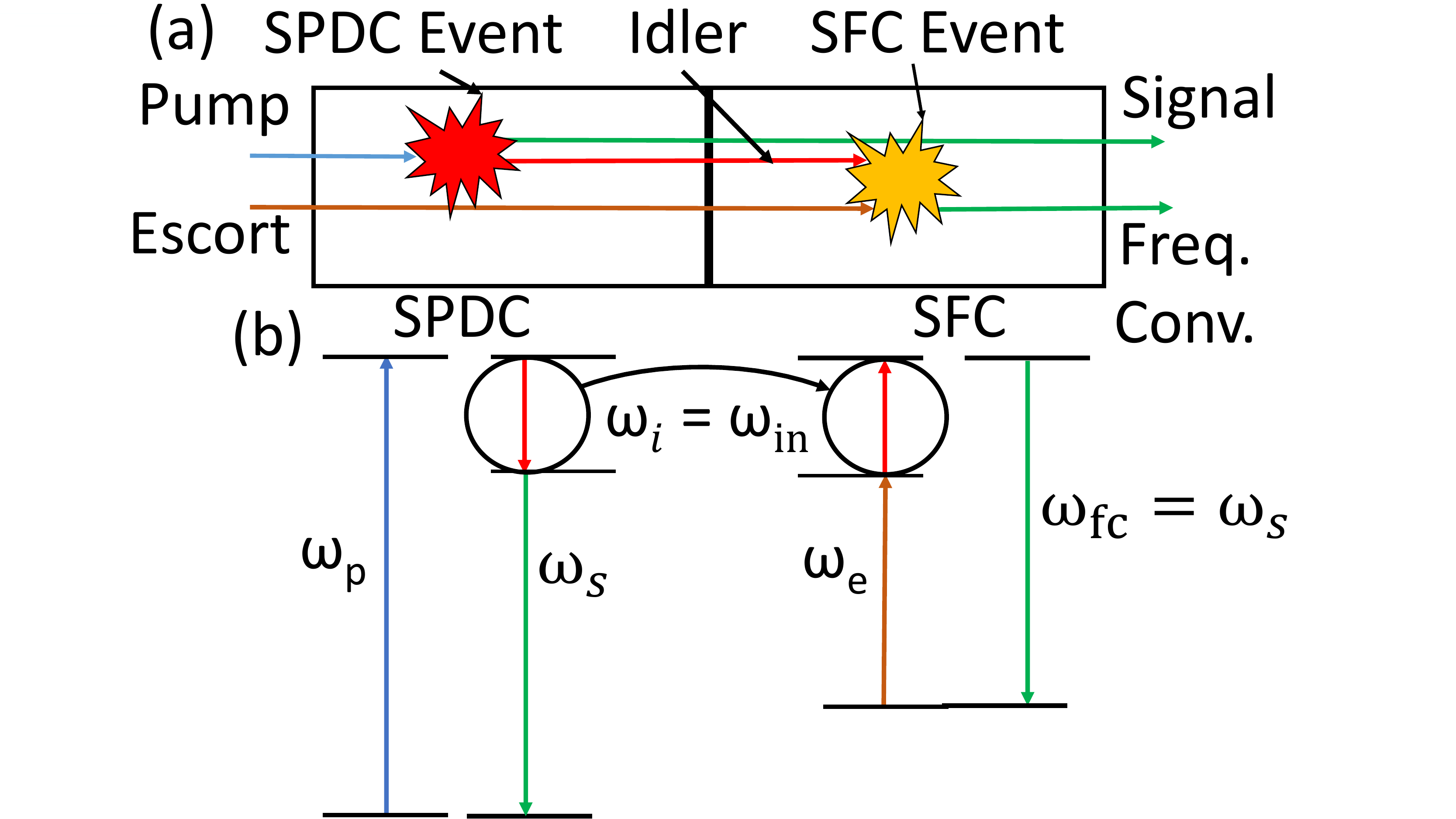}	
\caption{
(a) Depiction of our frequency-converted SPDC process, which can resolve the wavelength-purity trade-offs discussed in Sec. \ref{sec: Constraints SPDC} and thereby establish ultrahigh purity over a large spectral range.  
The nonlinear crystal is divided into two regions, where one region is poled for SPDC phase matching and the other for SFC phase matching.
The pulsed pump (blue) drives nondegenerate SPDC in the SPDC region. 
The idler photons from SPDC (red) are then frequency converted in the SFC region into the signal mode (green) by means of the temporally-coherent-high-power pulsed escort beam (brown).  
In (b) we give the energy diagram of the sequential SPDC and SFC processes; the idler ($\omega_i$) and escort photons ($\omega_e$) are combined to generate the frequency-converted photon ($\omega_{\mathrm{FC}}$), which has the same frequency as the signal photon ($\omega_s$).  In Fig. \ref{fig Visual Lambda Space} we show how this process can achieve ultrahigh purity.
}
\label{fig Freq Conv Diagrams}
\end{figure}

In this section we derive the output of our FC-SPDC scheme and discuss how it can generate photon pairs with high purity, and high indistinguishability.  
The basis of our scheme is to divide the nonlinear crystal into two regions where the poling structure and phase-matching conditions are first configured for SPDC and then sum-frequency conversion (SFC) [see Fig.~\ref{fig Freq Conv Diagrams}(a)].
Within the SPDC region, the phase matching is tuned such that a photon from a pump pulse decays into a nondegenerate signal and idler photon pair.
The idler photon then couples with an escort photon in the SFC region.
In a SFC process a photon at frequency $\omega_e$ from a high-powered pump, often called the escort, combines with a photon at frequency $\omega_{\mathrm{in}}$ and generates a higher-energy photon at the sum-frequency, which we simply call the frequency-converted photon $\omega_{\mathrm{FC}}$ \cite{brecht2011quantum,christ2013theory}.  
Our technique is to phase match the SFC region to establish degeneracy by converting the idler photon $\omega_{\mathrm{in}}=\omega_i$ to the signal frequency, thus $\omega_{\mathrm{FC}}=\omega_s$.
We assume the high-powered escort pulses are engineered to be temporally synchronized with the pulsed pump, and thus the escort pulses arrive on average in the SFC region at the same time as the idler photons (see Sec.~\ref{sec: entangled fc SPDC} and Fig.~\ref{fig Schematic Setup} for details.)
This ensures efficient SFC of escort and idler photons into the same frequency mode as the signal photon.
Next, we derive the output state of the device $\ket{\Psi_{\mathrm{out}}}$ and show how it can be engineered to have high purity and indistinguishability over a wide range of degeneracy wavelengths $\lambda_{\mathrm{deg}}$.

Analogous to SPDC, SFC can be modeled with the unitary operator~\cite{christ2013theory}
\begin{equation}
\label{U_SFG}
U_{\mathrm{SFC}} = \int\int f_{\mathrm{JCA}}(\omega_{\mathrm{in}},\omega_{\mathrm{FC}})\hat{a}_l(\omega_{\mathrm{in}})\hat{a}_m^\dagger(\omega_{\mathrm{FC}}) d\omega_{\mathrm{in}}d\omega_{\mathrm{FC}},
\end{equation}
where $l$ and $m$ are the polarizations of the input photon $\omega_{\mathrm{in}}$, which is the idler photon in our scheme, and the frequency-converted photon $\omega_{\mathrm{FC}}$, respectively.  
The spectral correlations between the input and frequency-converted photon are characterized by the joint conversion amplitude (JCA)
\begin{equation}
\label{equ: JCA}
f_{\mathrm{JCA}}(\omega_{\mathrm{in}},\omega_{\mathrm{FC}}) = \beta(\omega_{\mathrm{FC}}-\omega_{\mathrm{in}})\Psi(\omega_{\mathrm{in}},\omega_{\mathrm{FC}}),
\end{equation}  
where $\beta$ is the escort envelope function (EEF) and $\Psi$ is the PMF for SFC.  
One should note that energy conservation requires the EEF $\beta$ to be diagonal rather than antidiagonal.  
Therefore, $\alpha$ and $\beta$ are always orthogonal.  
The output state of the device is 
\begin{equation}
\label{equ: Psi out}
\ket{\Psi_{\mathrm{out}}} = U_{\mathrm{SFC}}U_{\mathrm{SPDC}}\ket{0}.
\end{equation}
In our scheme the SPDC idler photon is frequency converted, and thus we have the relation 
\begin{equation}
\label{equ: Commutator}
\hat{a}_l(\omega_{\mathrm{in}}) \hat{a}_j^\dagger(\omega_{i}) \ket{0}=\delta(\omega_i-\omega_{\mathrm{in}})\delta_{lj}\ket{0}.
\end{equation}
Substituting Eqs. (\ref{equ: JSA}), (\ref{equ: JCA}), and (\ref{equ: Commutator}) into Eq. (\ref{equ: Psi out}) and integrating over $\omega_{\mathrm{in}}$:
\begin{equation}
\label{equ: Psi out 2}
\ket{\Psi_{\mathrm{out}}}\!=\!\!\int\!\!f_{\mathrm{eff}}(\omega_{s},\omega_{\mathrm{FC}})\hat{a}_m^\dagger(\omega_{s})\hat{a}_k^\dagger(\omega_{\mathrm{FC}}) d\omega_{\mathrm{FC}}d\omega_s\ket{0},
\end{equation}
where we have defined the effective JSA
\begin{equation}
\label{JSA_eff}
f_{\mathrm{eff}}(\omega_{s},\omega_{\mathrm{FC}}) = \int\!\! f_{\mathrm{JCA}}(\omega_i,\omega_{\mathrm{FC}})f_{\mathrm{JSA}}(\omega_s,\omega_i)d\omega_i,
\end{equation}
which describes the spectral relationship of the two entangled photons leaving the device; the signal photon from the SPDC process and the frequency-converted photon from the SFC process.

As discussed in Sec.~\ref{sec: Constraints SPDC}, the purity $P$ and the indistinguishability $I$ of a biphoton state can be calculated from the shape of the JSA $f_{\mathrm{JSA}}(\omega_s,\omega_i)$.
Based on  Eq. (\ref{equ: JSA}), one can see that ultimately the shape of the JSA is determined by the orientation of the PMF $\Phi$ and the bandwidths of both the PEF $\alpha$ and the PMF $\Phi$.
Since the orientation of $\Phi$ is fixed for a given interaction, the bandwidths of $\alpha$ and $\Phi$ are the only free parameters that can be used to tune the state. 
The lack of tunability in general prevents the engineering of pure and indistinguishable output states at arbitrary $\lambda_{\mathrm{deg}}$.
Therefore, these sources are often designed with $\alpha$ and $\Phi$ of comparable bandwidths.
This is in contrast to our FC-SPDC scheme, which has more tunability as a consequence of the phenomenon contributing to the effective JSA $f_{\mathrm{eff}}$.
Based on Eq. (\ref{JSA_eff}), one can see that there are now four available bandwidth parameters; the bandwidths associated with the PMFs $\Phi$ and $\Psi$ and the envelope functions $\alpha$ and $\beta$.
Consequently, the problem becomes a four-parameter-optimization problem where one attempts to judiciously choose bandwidths that establish a two-dimensional (2D) Gaussian effective JSA.
We show in Sec.~\ref{sec: Results} that this is in general possible when one selects the optimal phase-matching configuration from Table \ref{tab: Config KTP} at the degeneracy wavelength $\lambda_{\mathrm{deg}}$ of interest.

\begin{figure}[t!]
\begin{tabular}{c|c}
	SPDC & SFC\\
	\hline
	\includegraphics[width=0.48\linewidth]{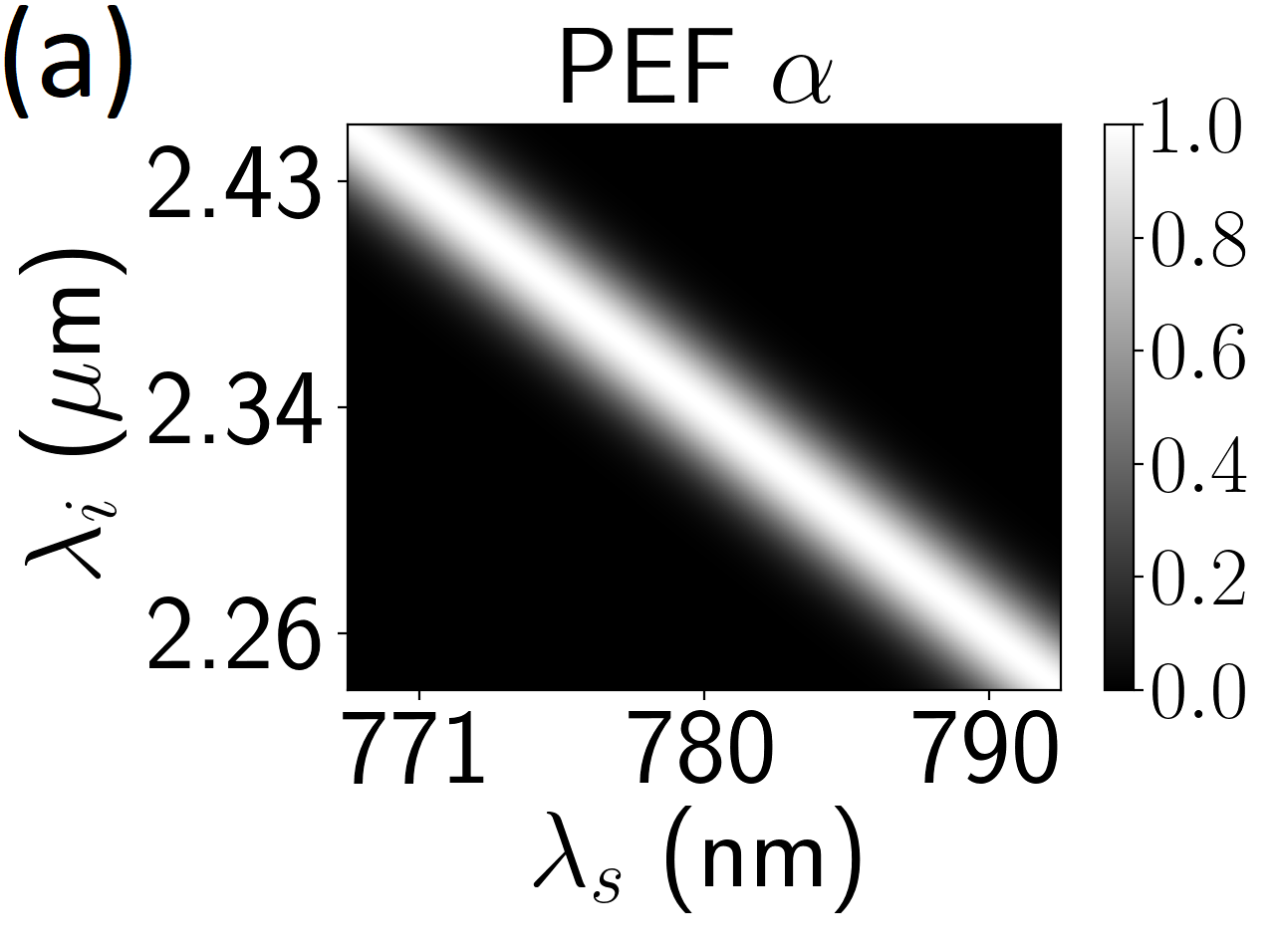} &
	\includegraphics[width=0.48\linewidth]{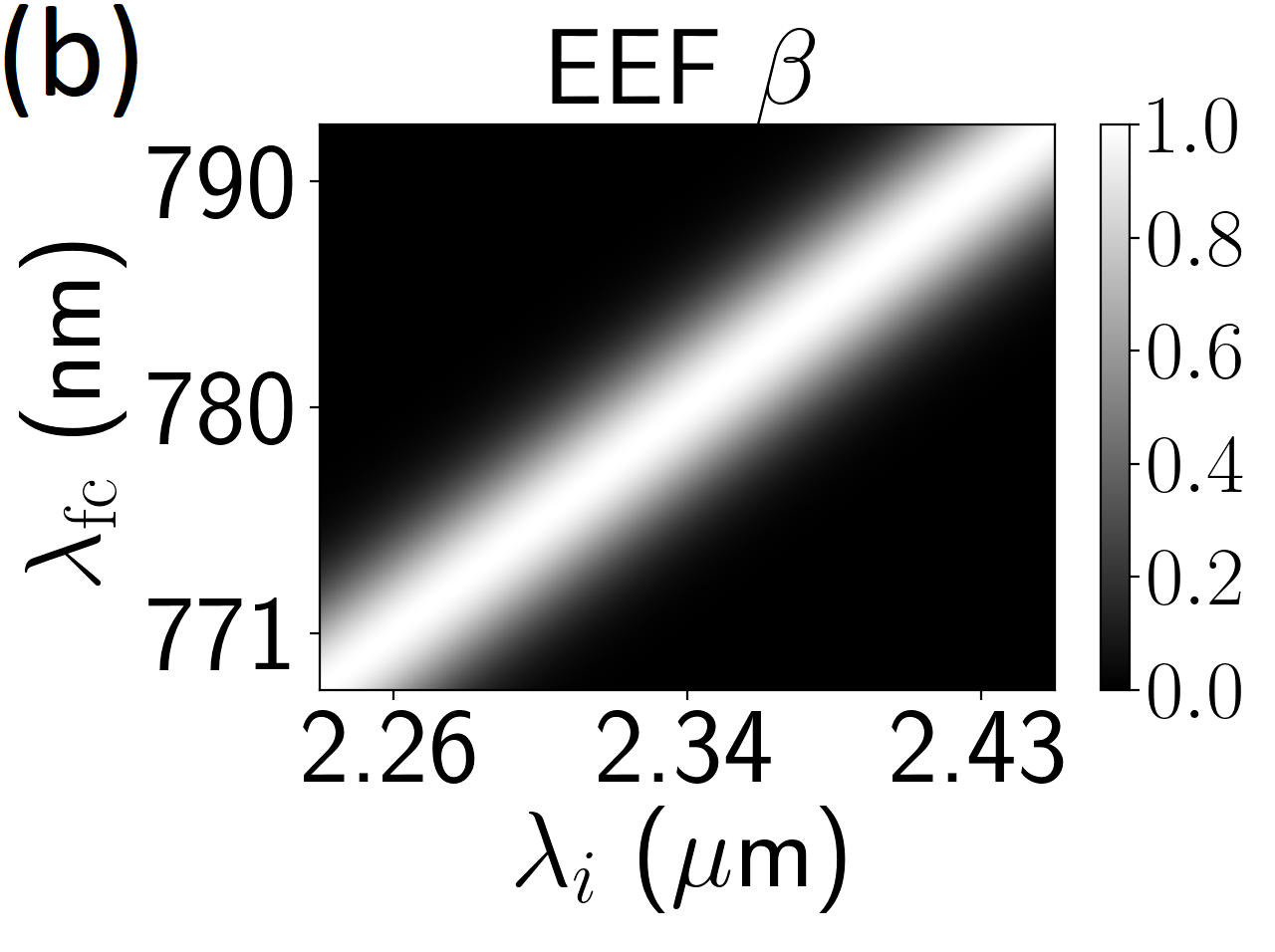}\\
	$\times$ & $\times$\\
	\includegraphics[width=0.48\linewidth]{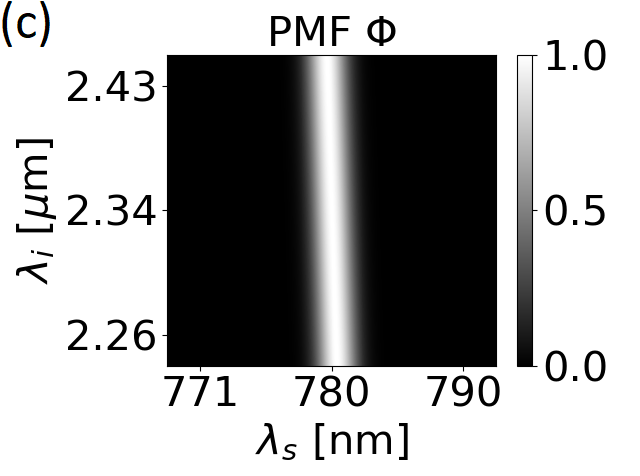} &
	\includegraphics[width=0.48\linewidth]{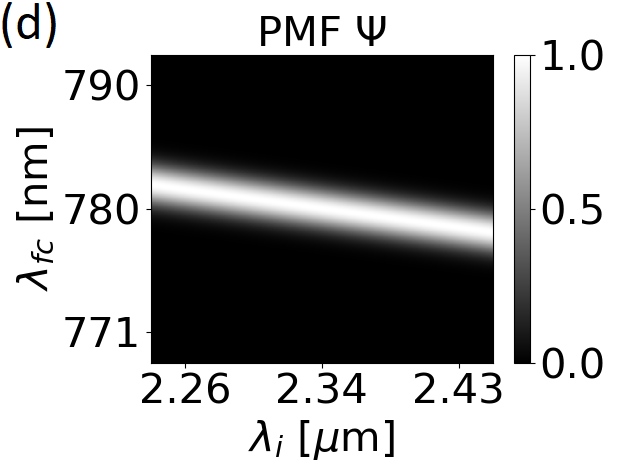}\\
		$=$ & $=$\\		
	\includegraphics[width=0.48\linewidth]{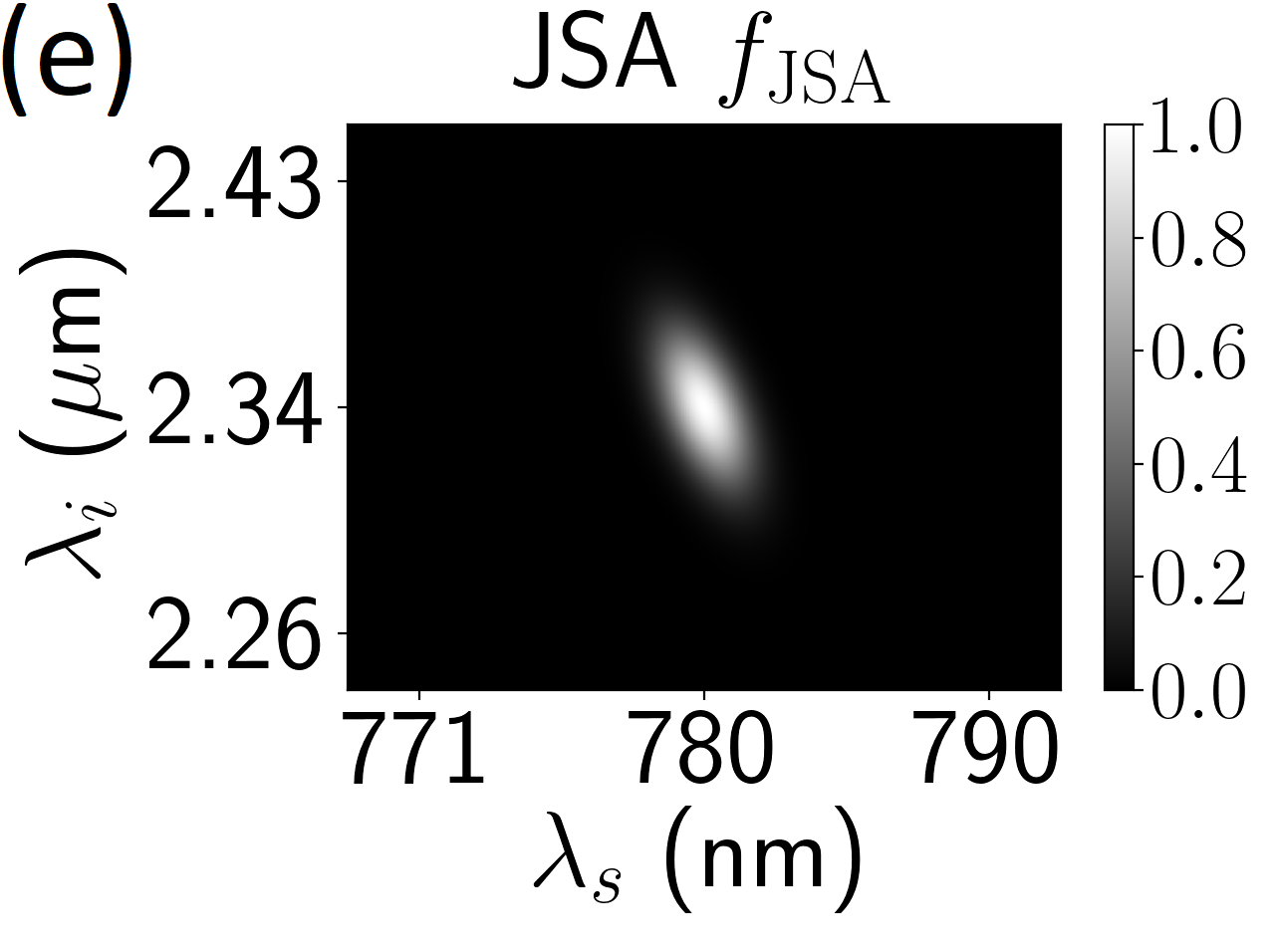} &

	\includegraphics[width=0.48\linewidth]{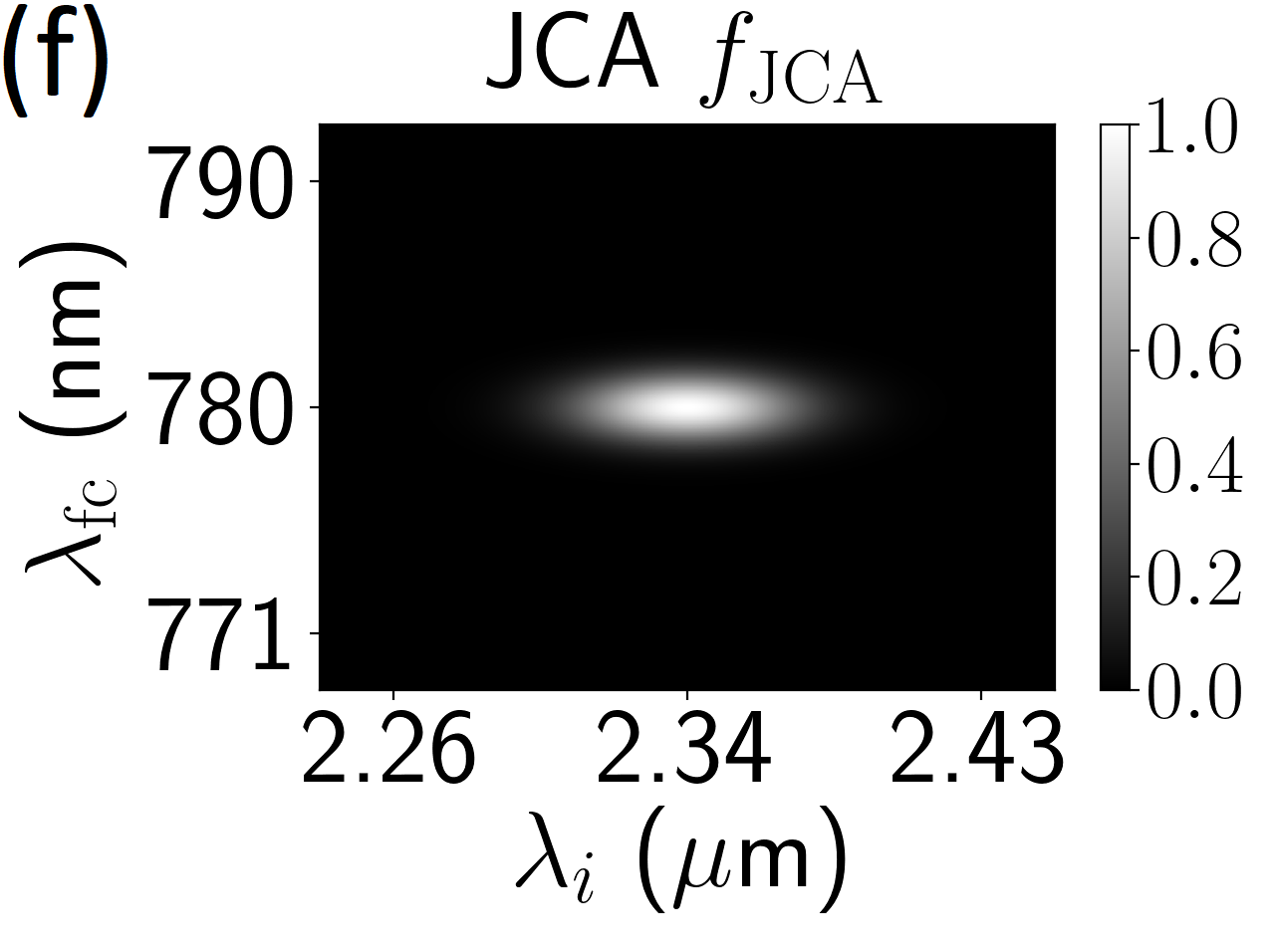}\\
\end{tabular}	
\begin{tabular}{c}
	\includegraphics[width=0.48\linewidth]{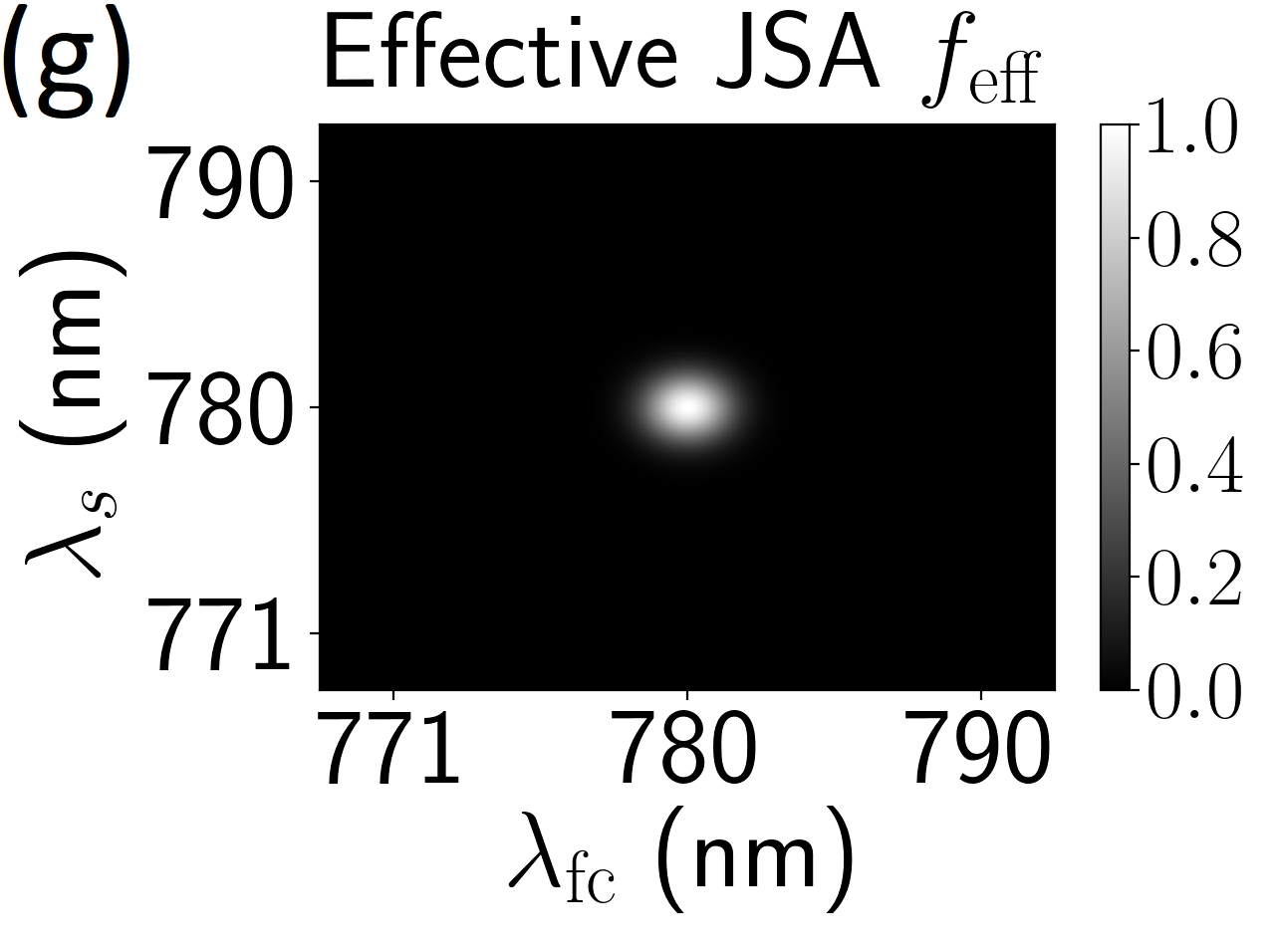}\\
\end{tabular}
\caption{Wavelength-space depiction of the FC-SPDC processes, which are depicted in Fig. \ref{fig Freq Conv Diagrams}, using the phase-matching configuration II from Table \ref{tab: Config KTP} at the degeneracy wavelength $\lambda_{\mathrm{deg}} = 780$ nm.  (a), (b) Antidiagonal PEF $\alpha$ and diagonal EEF $\beta$ represent energy conservation for the SPDC and SFC processes, respectively. (c), (d) PMFs $\Phi$ and $\Psi$ encode the crystal properties and momentum conservation. (e) JSA $f_{\mathrm{JSA}} = \alpha \times \Phi$ characterizes the spectral properties of the SPDC process. (f) JCA $f_{\mathrm{JCA}} = \beta \times \Psi$ characterizes the spectral properties of the SFC process.  (g) Effective JSA given by the matrix product $f_{\mathrm{eff}} = f_{\mathrm{JCA}} \times f_{\mathrm{JSA}}$ characterizes the output state. }
\label{fig Visual Lambda Space}
\end{figure}

As an example of this optimization process, Fig.~\ref{fig Visual Lambda Space} illustrates how to create a circular effective JSA at $\lambda_{\mathrm{deg}} = 780$ nm.
In this case, we utilize FC-SPDC in KTP with Gaussian phase matching according to the phase-matching configuration II in Table ~\ref{tab: Config KTP}.  
The left and right columns show the individual SPDC and SFC processes, respectively.  
Figures~\ref{fig Visual Lambda Space}(a), \ref{fig Visual Lambda Space}(c), and \ref{fig Visual Lambda Space}(e) give the PEF $\alpha$, PMF $\Phi$, and JSA $f_{\mathrm{JSA}}$ for the SPDC process.
Therefore, the left column is an illustration of Eq. (\ref{equ: JSA}), that is, $f_{\mathrm{JSA}}$ is the scalar product of $\alpha$ and $\Phi$.
Similarly, Figs.~\ref{fig Visual Lambda Space}(b), \ref{fig Visual Lambda Space}(d), \ref{fig Visual Lambda Space}(f) are the EEF $\beta$, PMF $\Psi$, and JCA $f_{\mathrm{JCA}}$ for the SFC process, where $f_{\mathrm{JCA}}$ is the scalar product of $\beta$ and $\Psi$ as stated in Eq. (\ref{equ: JCA}).
The effective JSA $f_{\mathrm{eff}}$ is shown in Fig.~\ref{fig Visual Lambda Space}(g).
Considering Eq. (\ref{JSA_eff}) in discretized form, $f_{\mathrm{eff}}$ becomes the matrix product of $f_{\mathrm{JCA}}$ [Fig.~\ref{fig Visual Lambda Space}(f)] and $f_{\mathrm{JSA}}$ [Fig.~\ref{fig Visual Lambda Space}(e)].

\setlength{\arrayrulewidth}{0.1em}
The purity $P$ and the indistinguishablity $I$ are unity for this example, because at this $\lambda_{\mathrm{deg}}$ and phase-matching configuration the orientations of $\Phi$ and $\Psi$ reside within favorable regions of the parameter space.
Consequently, one can easily identify bandwidth parameters that result in a circular Gaussian $f_{\mathrm{eff}}$.
However, for certain orientations of $\Phi$ and $\Psi$ this is not possible.
Therefore, we establish an optimization procedure that selects bandwidths that maximize the quantity $\eta=P \times I$.
Accordingly, our method is to search all the possible phase matching configurations (listed in Table ~\ref{tab: Config KTP}) and to select the configuration with the largest $\eta$ for the $\lambda_{\mathrm{deg}}$ of interest.
We will discuss this further in Sec.~\ref{sec: Results}.

\section{Entangled Frequency-converted SPDC Source}
\label{sec: entangled fc SPDC}
In this section we discuss a method to generate polarization entanglement using our FC-SPDC process and a Sagnac interferometer.
The entanglement is ideal in the sense that there are no detrimental frequency-polarization correlations \cite{poh2007joint,poh2009eliminating}.

\begin{figure}[t!]
\includegraphics[width=1\linewidth]{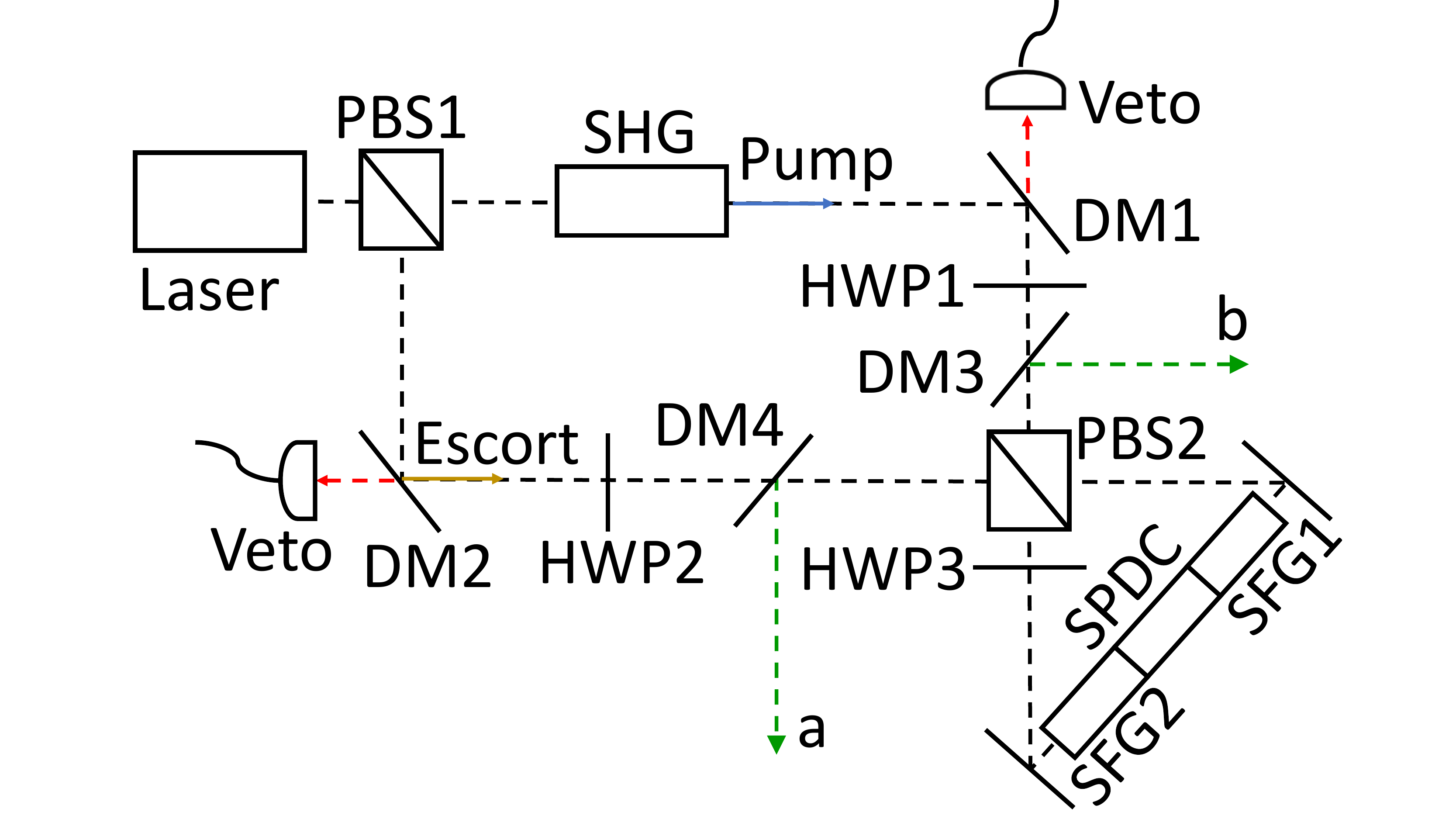}
\caption{Configuration for generating polarization entanglement using our FC-SPDC scheme.  A Mach-Zehnder interferometer prepares the pump and escort pulses: the laser, which is configured at the escort pump frequency, is split at the polarizing beam splitter (PBS1),  in the upper path the second-harmonic generation (SHG) crystal generates the pump beam, the polarizations of the beams are rotated to $45^\circ$ with the half-wave plates (HWP1 and HWP2), and the beams are recombined at polarizing beam splitter PBS2.  The two-region nonlinear crystal depicted in Fig.~\ref{fig Freq Conv Diagrams}(a) is modified to contain a SPDC region placed between two SFC regions, making the crystal symmetric about the center.  This three-region crystal is placed at the center of a Sagnac interferometer integrated with a half-wave plate (HWP3) to generate identical frequency-converted SPDC photons in both the clockwise and counterclockwise directions.  As a result the output modes of the polarizing beam splitter (PBS2) are reflected at the dichroic mirrors (DMs), and thus the reflected modes a and b are polarization entangled. A detection at $\lambda_i$ in a ``Veto" mode allows one to maintain high heralding efficiency by detecting when the SFC process failed and vetoing that pulse.}
\label{fig Schematic Setup}
\end{figure}

A schematic of our method is shown in Fig.~\ref{fig Schematic Setup}.  
The preliminary step is to generate temporally synchronized escort and pump pulsed beams that will separately drive the SFC and SPDC processes, respectively.  
A configuration that accomplishes this is shown in the upper (Mach-Zehnder) portion of Fig.~\ref{fig Schematic Setup}.   
A high-powered pulsed laser configured at the escort frequency $\omega_e$ is divided at the polarizing beam splitter PBS1.  
The beam in the upper path is frequency doubled in a nonlinear second-harmonic generation (SHG) crystal, generating the pump beam at frequency $\omega_p=2\omega_e$.  
The linear polarizations of the pump beam and the escort beam (lower path) are set to $45^\circ$ by the half-wave plates HWP1 and HWP2.  
Lastly, the pump and escort beams are recombined at the PBS2.  
By deriving the escort and the pump beams from the same high-powered pulsed laser, the temporal offset  between the escort and pump pulses can be tuned and held fixed so that the idler photons created in the SPDC region arrive on average in the center of the SFC region at the same time as the escort pulses. 

For the next step, the escort and pump beams enter the Sagnac loop at the polarizing beam splitter PBS2 and are each equally divided into clockwise and counterclockwise traveling pulses \cite{shi2004generation,predojevic2012pulsed}.  
Centered in the loop is the nonlinear crystal, which is symmetrized with respect to the center of the loop so that identical frequency-conversion processes occur for both propagation directions.
To achieve this symmetry we place the SPDC region of the crystal between two identical SFC regions.
Due to the quasi-phase-matching, the pump and the escort pulses interact almost exclusively within the SPDC region and the SFC region, respectively. 
Therefore, unwanted nonlinear interactions are suppressed.
Spurious nonlinear interactions that persist should generate photons outside the wavelength band of interest, and thus should be easy to filter out.

Utilizing a Sagnac loop to generate polarization entanglement requires the frequency-converted photon and the signal photon to have orthogonal polarizations, otherwise they will not separate at PBS2 into different spatial modes.  
Furthermore, the inclusion of the half wave plate HWP3 ensures that the signal photons always exit PBS2 into the same mode regardless of the propagation direction.
Similarly, the frequency-converted photons always exit PBS2 into the other mode.
In this way the output state at spatial modes $a$ and $b$ has no frequency-polarization correlations.
We call this ideal polarization entanglement, because any distinguishing spectral information contained in the output state has no effect on the quality of the polarization entanglement.
The following frequency relationships result from our scheme: 
\begin{equation}
\begin{split}
\omega_p&=2\omega_e, \; \omega_{\mathrm{FC}}=\omega_s, \; \mathrm{and}\; \omega_i=\frac{\omega_s}{3}.
\end{split}
\end{equation}

At the end of Sec.~\ref{sec: Constraints SPDC} we noted that our scheme can be interpreted as a more elegant filtering strategy.
This should be evident from the the expression for the effective JSA in Eq. (\ref{JSA_eff}) and the progression leading to the circular effective JSA in Fig.~\ref{fig Visual Lambda Space}.
A consequence of this filtering effect is the SFC conversion efficiency which can be defined as the ratio of the joint spectral intensities (JSIs) 
\begin{equation}\label{eq:conv_eff}
\begin{split}
\eta_{\mathrm{conv}} \equiv  | f_{\mathrm{eff}}| ^2 /  | f_{\mathrm{JSA}}| ^2.
\end{split}
\end{equation}
The conversion efficiency can be low for certain wavelength configurations, but in most cases it is not as significant as the attenuation that would be observed with conventional filtering (see Fig.~\ref{fig Conv Perc}).
Furthermore, whereas conventional filtering can significantly decrease heralding efficiency, our method can maintain high heralding efficiency by detecting unconverted idler photons and using this information to ``Veto" a pulse in which a degenerate-entangled pair was not created  (see red dashed lines labeled ``Veto" in Fig.~\ref{fig Schematic Setup}).
In practice there will be other contributions to the conversion efficiency.  
One such contribution is the incomplete conversion that occurs when the escort pump power is too low \cite{brecht2011quantum,christ2013theory}.  
This would be accounted for by a prefactor in Eq. (\ref{eq:conv_eff}).
However, since our scheme allows independent control of the escort beam, this contribution to the efficiency is ignored.  
Other inefficiencies could be introduced, for example, by limited transverse spatial mode overlap, which we discuss in the following paragraph.

The well-known problems related to spatial and transverse-spatial mode optimization inherent to conventional SPDC in crystals are likewise inherent to our method.  
Utilizing a bulk crystal, one must consider how to optimize beam overlap including how to focus the pump beam in order to maximize certain measures \cite{bennink2010optimal}.
This issue is compounded in our scheme due to the three phase-matching regions, but conventional solutions still apply. 
The main difference is that one must consider the challenge of focusing the pump at the center of the SPDC region and the escort at the center of the SFC regions.
Likewise, the transverse-spatial-mode overlap optimization must be performed for both the SPDC and SFC regions \cite{yang2008spontaneous, lanning2018quantized}.
Common to both methods, one must ask what performance measure to optimize for and how this relates to beam parameters.
For example, if one were to couple spatial modes $a$ and $b$ (see Fig.~\ref{fig Schematic Setup}) into single-mode fibers, then the beam parameters for optimal pair rate will likely be very different than the optimal parameters for heralding efficiency \cite{fedrizzi2007wavelength}.
This is a result of the higher-order transverse-spatial-mode structure of the SPDC photons.
The other option is to use a nonlinear waveguide approach, which can give up to a 50$\times$ improvement in down-conversion rate \cite{Fiorentino:07}.
However, the waveguides can also support SPDC in higher-order transverse-spatial modes due to the disparate wavelengths of the pump and down-converted photons, and due to imperfections in waveguide manufacturing.
Additionally, the problem can be compounded because the modes of the waveguide may have poor transverse-spatial mode overlap with the assumed fiber modes \cite{christ2009spatial,mosley2009direct}.
Therefore, a similar optimization problem is required. 
Ideally, one would couple all the wavelengths into their respective fundamental waveguide mode \textit{and} have strong overlap with single mode fiber, thus minimizing detrimental transverse spatial mode effects. 
Ultimately, the issue of transverse-spatial-mode optimization is application specific, and thus reserved for future work.

\section{Simulation Results}
\label{sec: Results}
In this section we present results that demonstrate the utility of FC-SPDC scheme.
Keeping the crystal temperature constant at 20$^{\circ}$ C, we span a large parameter space in which we find the best performance possible with KTP, and LN.
For each $\lambda_{\mathrm{deg}}$, we also consider the effects of both periodic poling and domain engineering, which result in sinc PMFs or ideal Gaussian PMFs, respectively.
Recall from Secs. \ref{sec: FC SPDC} and \ref{sec: entangled fc SPDC} that the idler photon is frequency converted in the SFC region, and to establish polarization entanglement in the Sagnac configuration the frequency-converted photon must be orthogonal to the signal photon.
These requirements limit the possible phase-matching configurations to those labeled I through VIII in Table \ref{tab: Config KTP}, where $x$, $y$, and $z$ represent photons polarized along the respective crystolographic axis (X,Y,Z), and SPDC and SFC indicate the nonlinear process.
For the case of the uniaxial crystal LN only the configurations I through IV are relevant.
Furthermore, the indices of refraction $n_y$ and $n_z$ become the ordinary and extraordinary indices, respectively.
The photons involved in the SPDC and SFC processes are labeled from left to right: pump $\rightarrow$ idler $+$ signal and escort $+$ idler $\rightarrow$ frequency converted, respectively.

\begin{table}[b!]
\caption{The phase-matching configurations where $x$, $y$, $z$ are photons polarized along the respective crystolographic axis (X, Y, Z).  The columns for SPDC are pump $\rightarrow$ idler $+$ signal, and the columns for SFC are escort $+$ idler $\rightarrow$ frequency converted.  For the nonlinear crystal LN, $n_y$ and $n_z$ are the ordinary and extraordinary refractive indices, respectively.}
\resizebox{0.8\columnwidth}{!}{
\begin{tabular}{|l|c|r|}
\hline
Conf. Num. & SPDC & SFC \\
\hline
I & $y \rightarrow y+z$ & $z+y \rightarrow y$ \\
II & $y \rightarrow z+y$ & $z+z \rightarrow z$ \\
III & $z \rightarrow z+z$ & $y+z \rightarrow y$ \\
IV & $z \rightarrow y+y$ & $y+y \rightarrow z$ \\
V & $x \rightarrow x+z$ & $z+x \rightarrow x$ \\
VI & $x \rightarrow z+x$ & $z+z \rightarrow z$ \\
VII & $z \rightarrow z+z$ & $x+z \rightarrow x$ \\
VIII & $z \rightarrow x+x$ & $x+x \rightarrow z$ \\
\hline
\end{tabular}
}
\label{tab: Config KTP}
\end{table}

The results we present proceed as follows.
For each $\lambda_{\mathrm{deg}}$, we choose the configuration in Table ~\ref{tab: Config KTP} that maximizes $\eta=P\times I$.
We also evaluate our periodically poled source configuration in terms of heralding efficiency $H$ after filtering the minimum necessary to remove the sinc sidebands from $f_\mathrm{eff}$:
\begin{equation}
H = \frac{\mathcal{P}_{\mathrm{both}}}{\mathcal{P}_{i|j}},
\end{equation}
where $\mathcal{P}_{\mathrm{both}}$ is defined in Eq. (\ref{eq:P_BOTH}), and $\mathcal{P}_{i|j}$ is the conditional probability of detecting one photon with unit detection efficiency provided that the second of the pair passed through a filter and was detected in the other mode.
For reference, we compare these metrics to those of an equivalent conventional degenerate SPDC photon source. 
The simulations utilizing KTP use the Sellmeier equations from Refs. \cite{kato2002sellmeier, emanueli2003temperature}, and the simulations utilizing LN or MgLN use the Sellmeier equations from Ref. \cite{zelmon1997infrared}. 

\begin{figure*}
	\begin{subfigure}{0.45\linewidth}
	\includegraphics[width=\linewidth]{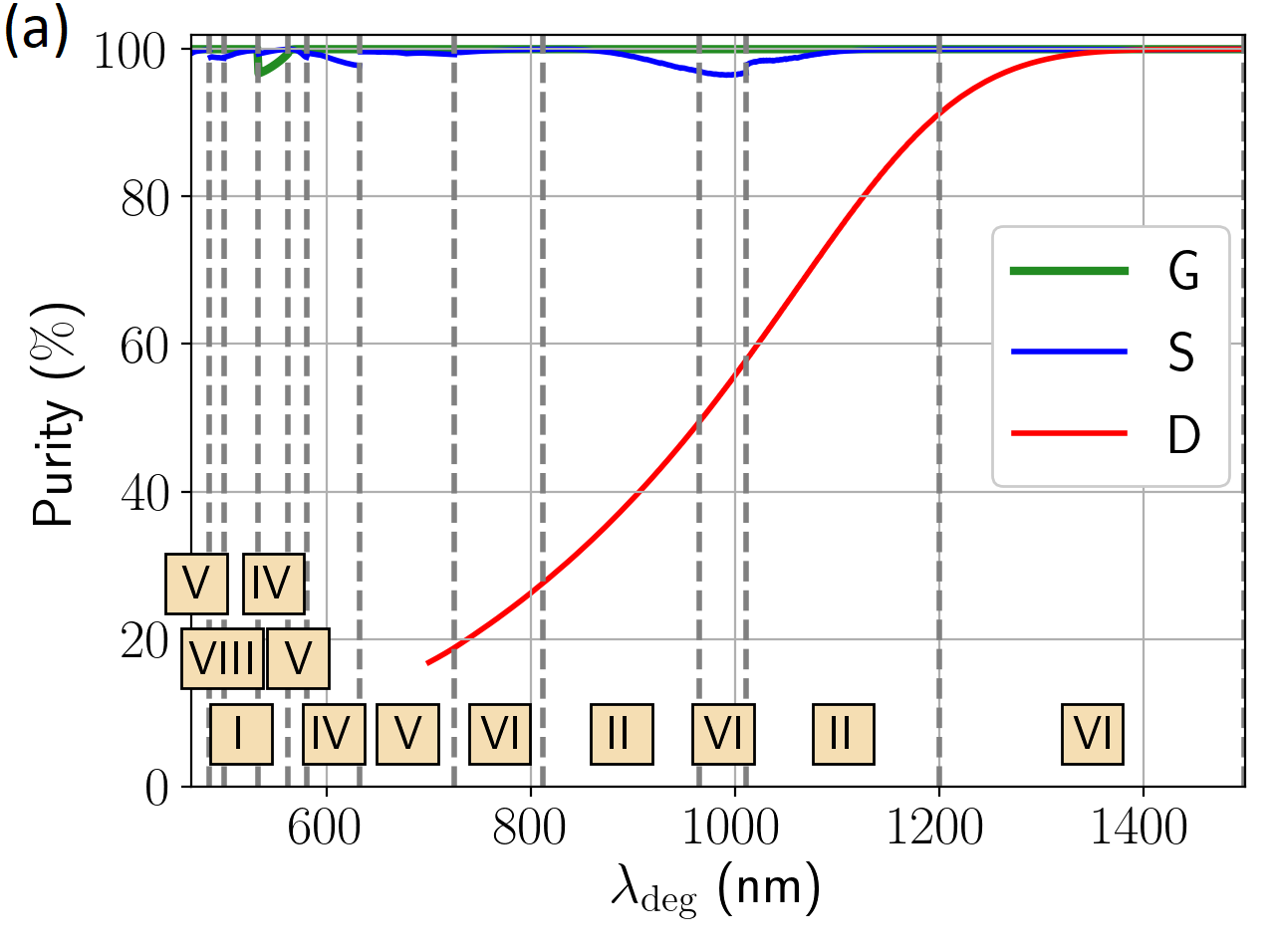}
	\label{fig P vs lam KTP}
	\end{subfigure}%
	\begin{subfigure}{0.45\linewidth}
	\includegraphics[width=\linewidth]{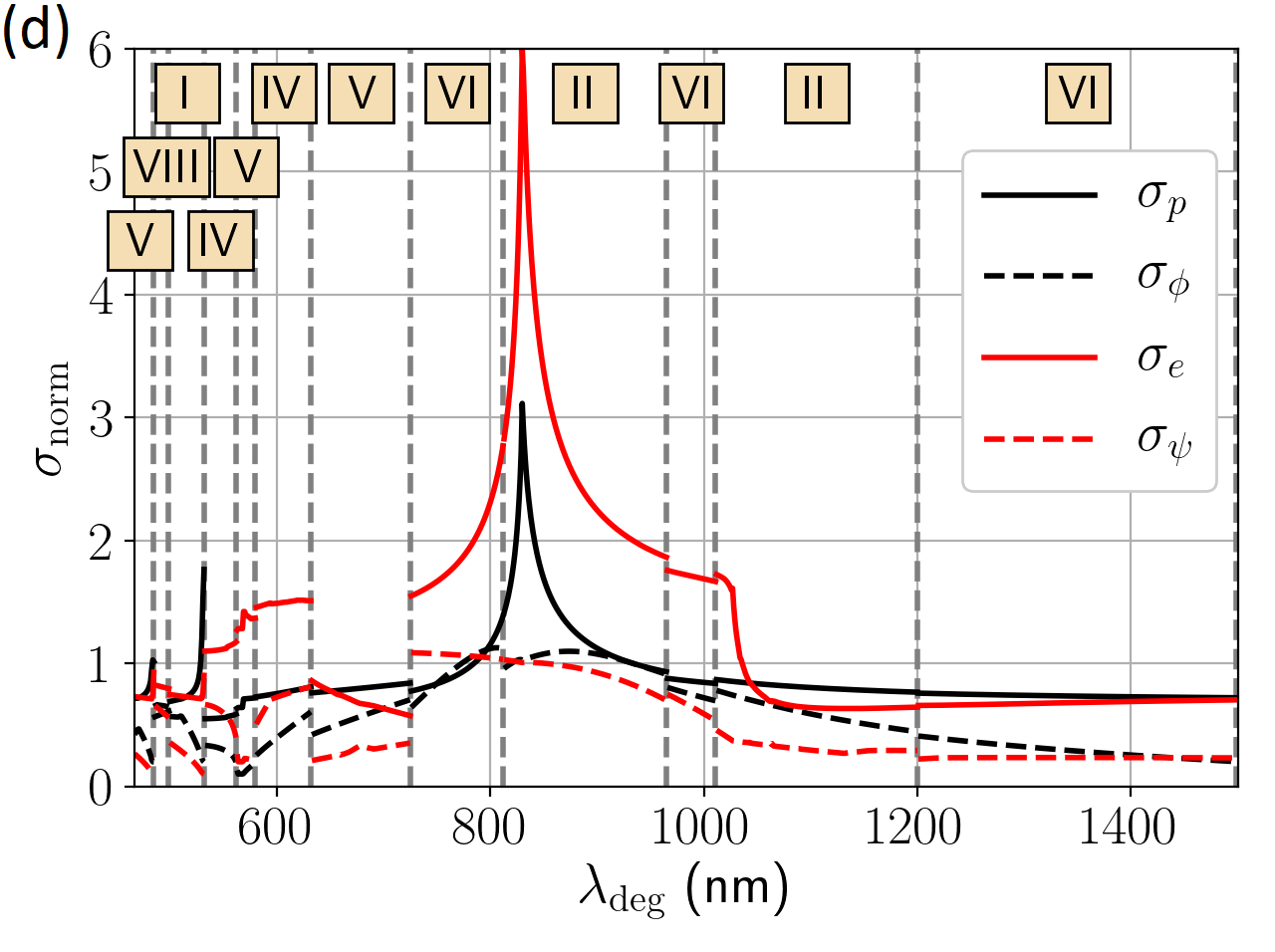}
	\label{fig sig norm vs lam KTP}
	\end{subfigure}
	\begin{subfigure}{0.45\linewidth}	
	\includegraphics[width=\linewidth]{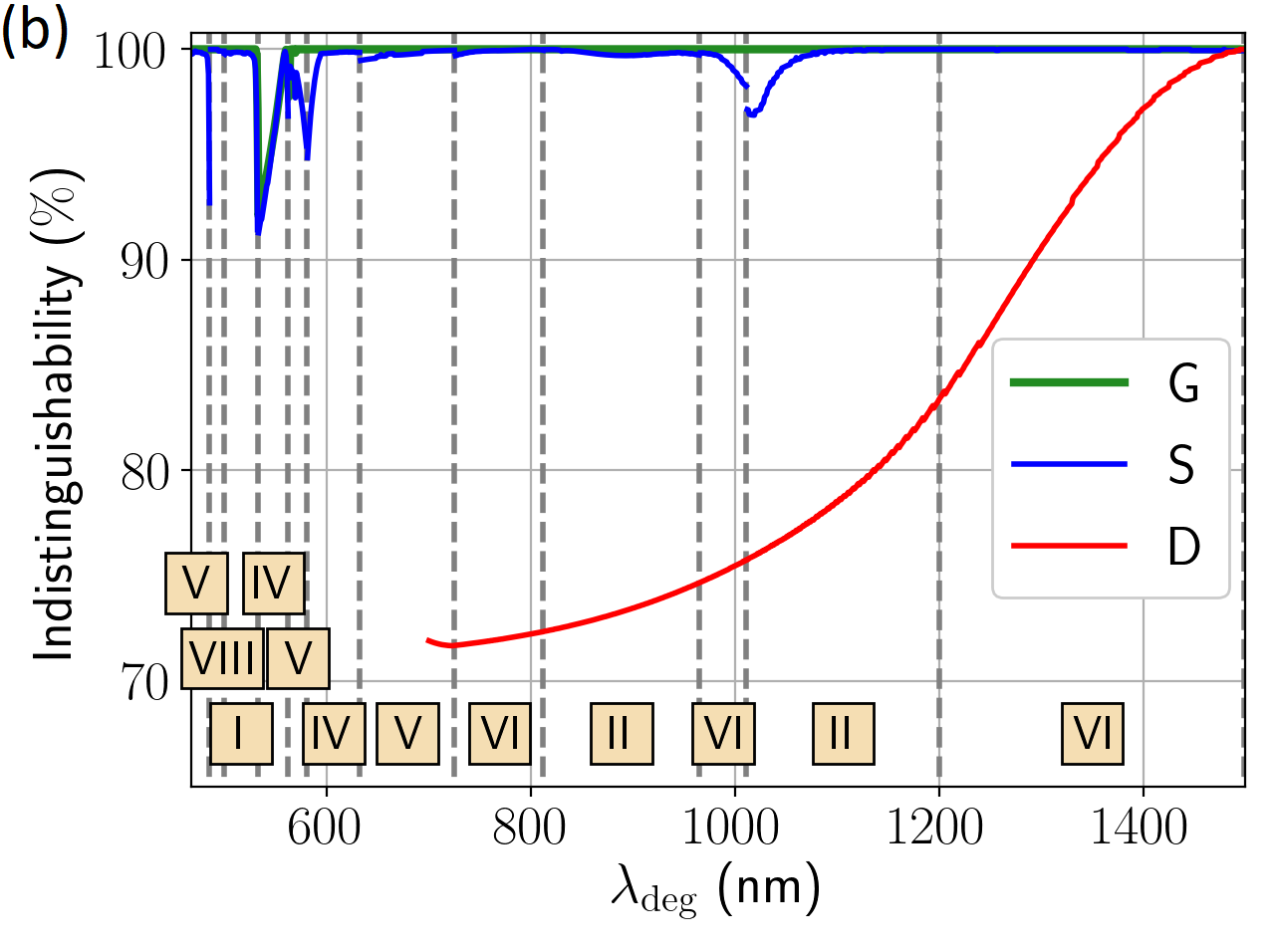}
	\label{fig I vs lam KTP}
	\end{subfigure}%
	\begin{subfigure}{0.45\linewidth}	
	\includegraphics[width=\linewidth]{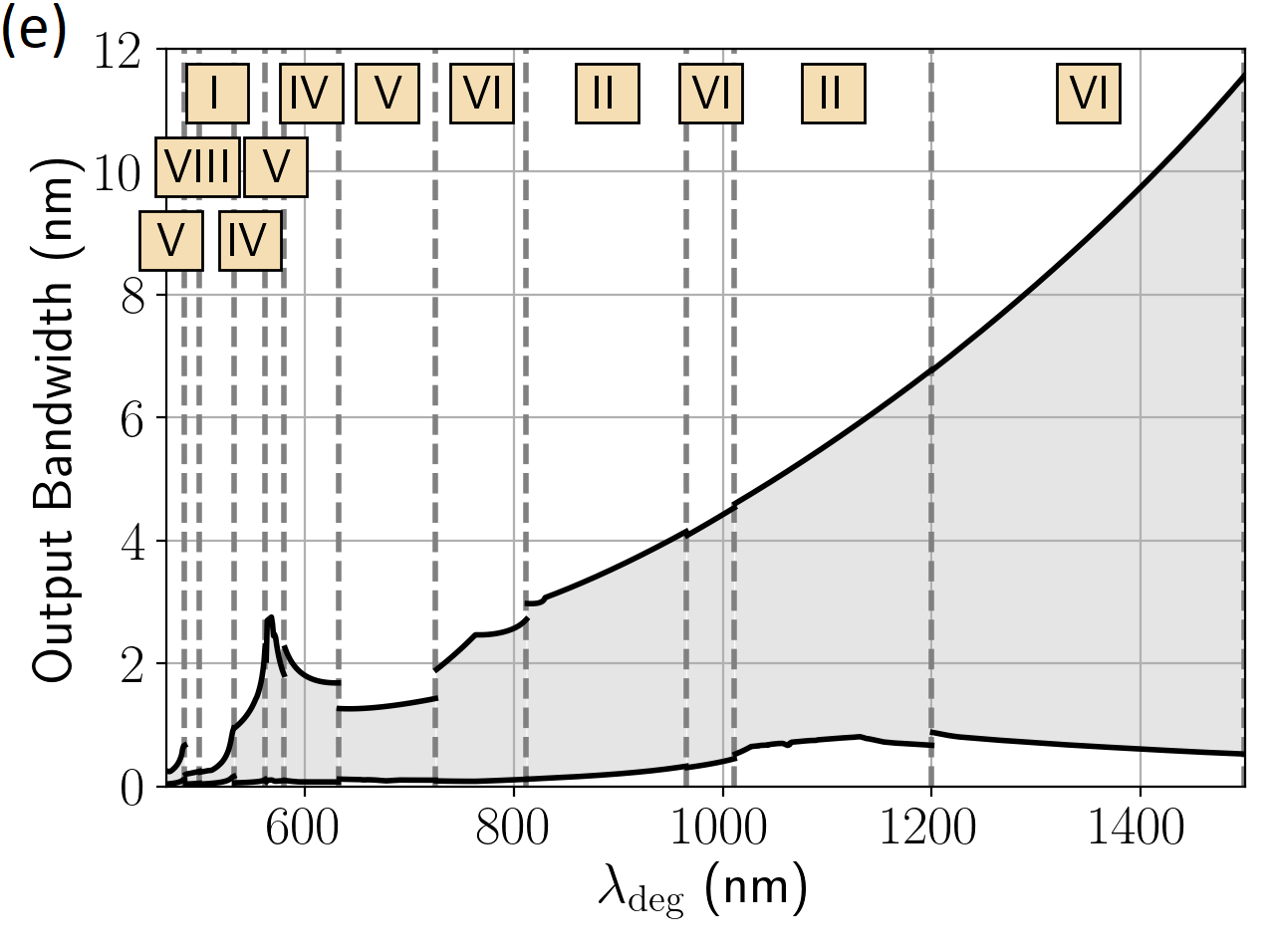}
	\label{fig sig out vs lam KTP}
	\end{subfigure}
	\begin{subfigure}{0.45\linewidth}	
	\includegraphics[width=\linewidth]{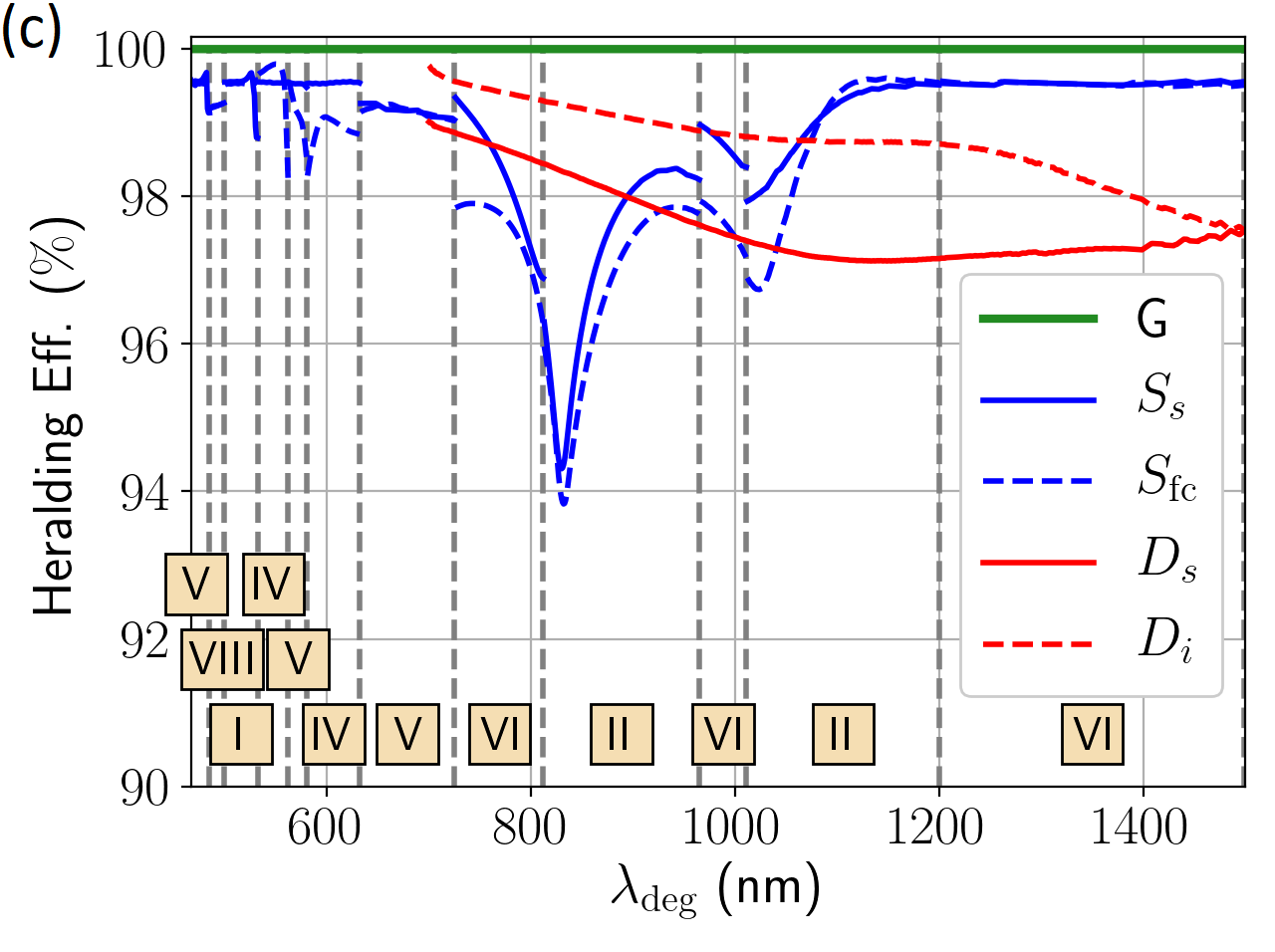}
	\label{fig H vs lam KTP}
	\end{subfigure}%
	\begin{subfigure}{0.45\linewidth}	
	\includegraphics[width=\linewidth]{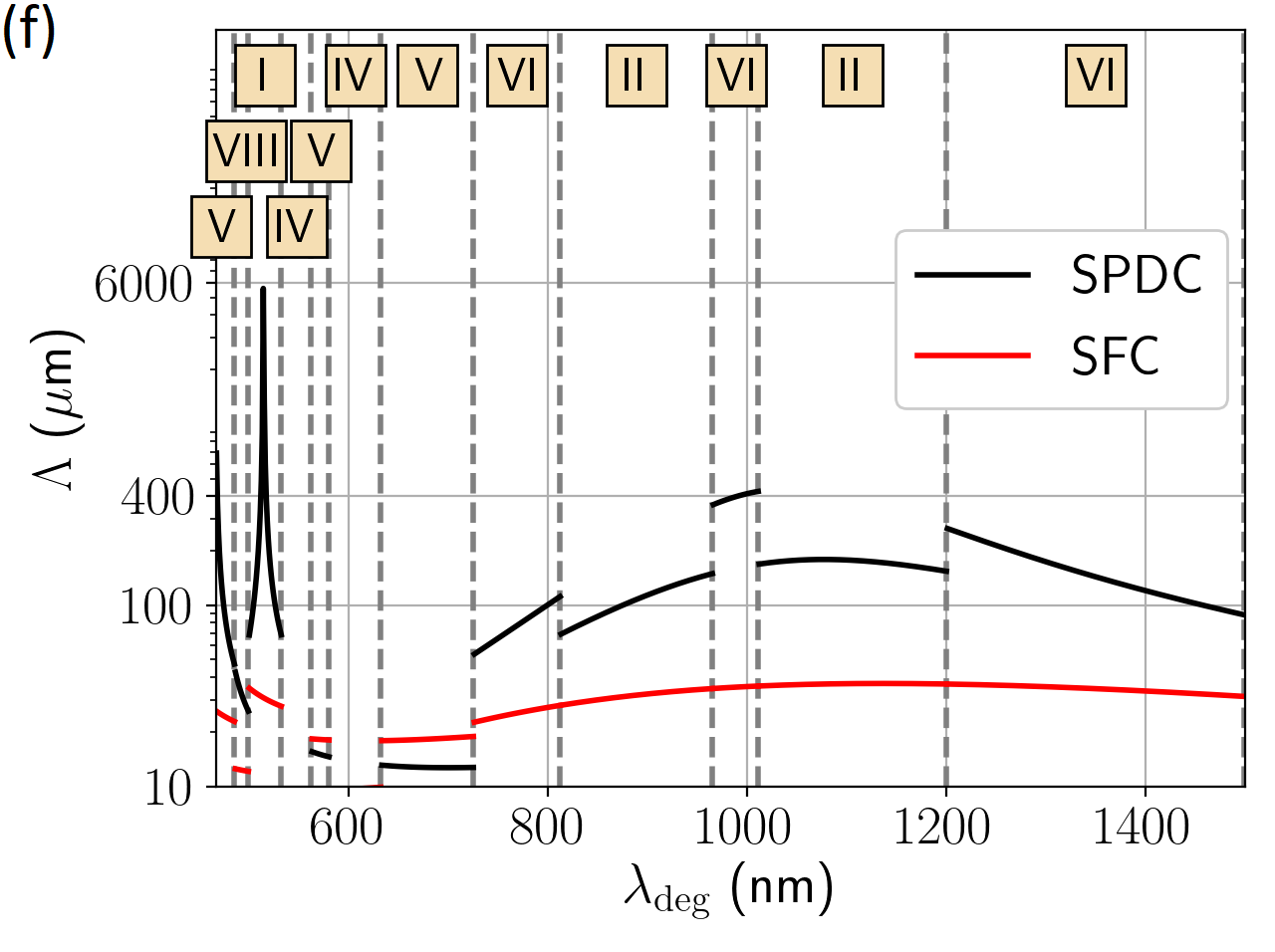}
	\label{fig Pol vs lam KTP}
	\end{subfigure}%
	\caption{The results of the frequency-converted SPDC optimization for the nonlinear crystal KTP. The figure includes the (a) purity, the (b) indistinguishability, and (c) the heralding efficiency for frequency-converted SPDC with Gaussian G and sinc S phase matching, and for conventional-degenerate SPDC D with sinc phase matching (the curves S and D are filtered just enough to remove the sidebands from the JSA), the (d) optimal pump ($\sigma_p$), escort ($\sigma_e$), SPDC phase-matching bandwidth ($\sigma_\phi$), and SFC phase-matching bandwidth ($\sigma_\psi$) normalized to the output bandwidth of the device, the (e) achievable output bandwidths, and the (f) poling period for the SPDC and SFG regions of the crystal.  The optimal phase-matching configurations at each output wavelength ($\lambda_{\mathrm{deg}}$) are labeled by Roman numerals corresponding to the rows in Table \ref{tab: Config KTP}.  The boundaries are indicated by the vertical gray dashed lines. 
}
	\label{fig Result KTP}
\end{figure*}

First, we present the results of our FC-SPDC scheme as a function of $\lambda_{\mathrm{deg}}$ using KTP (see Fig.~\ref{fig Result KTP}).
The metrics $P$, $I$, and $H$ are shown in Figs.~\ref{fig Result KTP}(a)--\ref{fig Result KTP}(c), respectively. 
The optimal bandwidth parameters are shown in Fig.~\ref{fig Result KTP}(d), the range of allowable output bandwidths are shown in Fig.~\ref{fig Result KTP}(e), and the required poling period is shown in Fig.~\ref{fig Result KTP}(f).    
The optimal phase-matching configuration for periodic poling and sinc phase matching is indicated by the Roman numeral I through VIII (see Table \ref{tab: Config KTP}) and the boundaries of each configuration by the vertical, gray lines.  
The green curves labeled G correspond to the FC-SPDC scheme with Gaussian phase matching, the blue curves labeled S correspond to the FC-SPDC scheme with sinc phase matching, and red curves labeled D in Figs.~\ref{fig Result KTP}(a)--\ref{fig Result KTP}(c) correspond to conventional degenerate SPDC with sinc phase matching.
We apply the constraint

\begin{figure*}
	\begin{subfigure}{0.45\linewidth}
	\includegraphics[width=\linewidth]{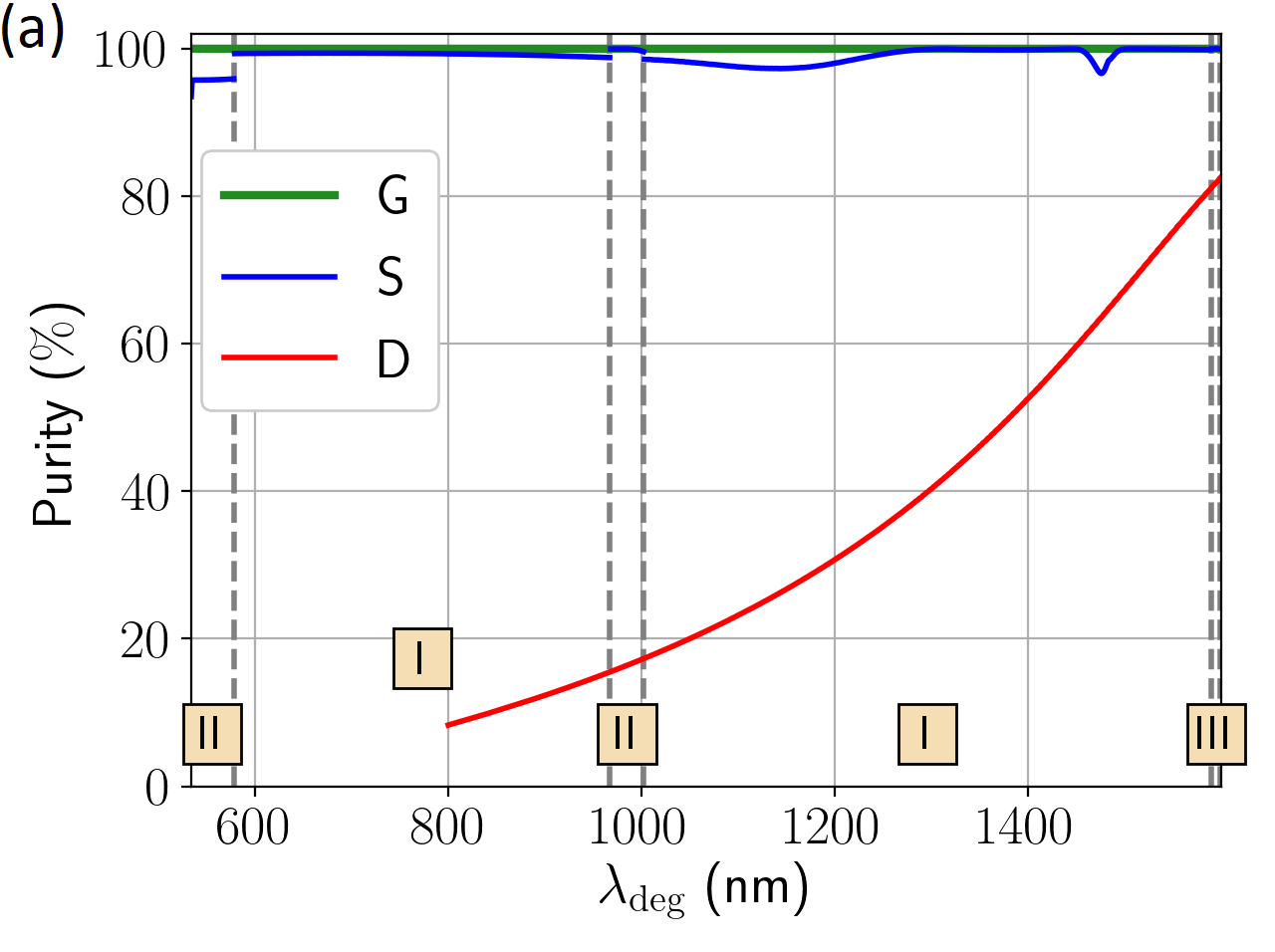}
	\label{fig P vs lam LN}
	\end{subfigure}%
	\begin{subfigure}{0.45\linewidth}
	\includegraphics[width=\linewidth]{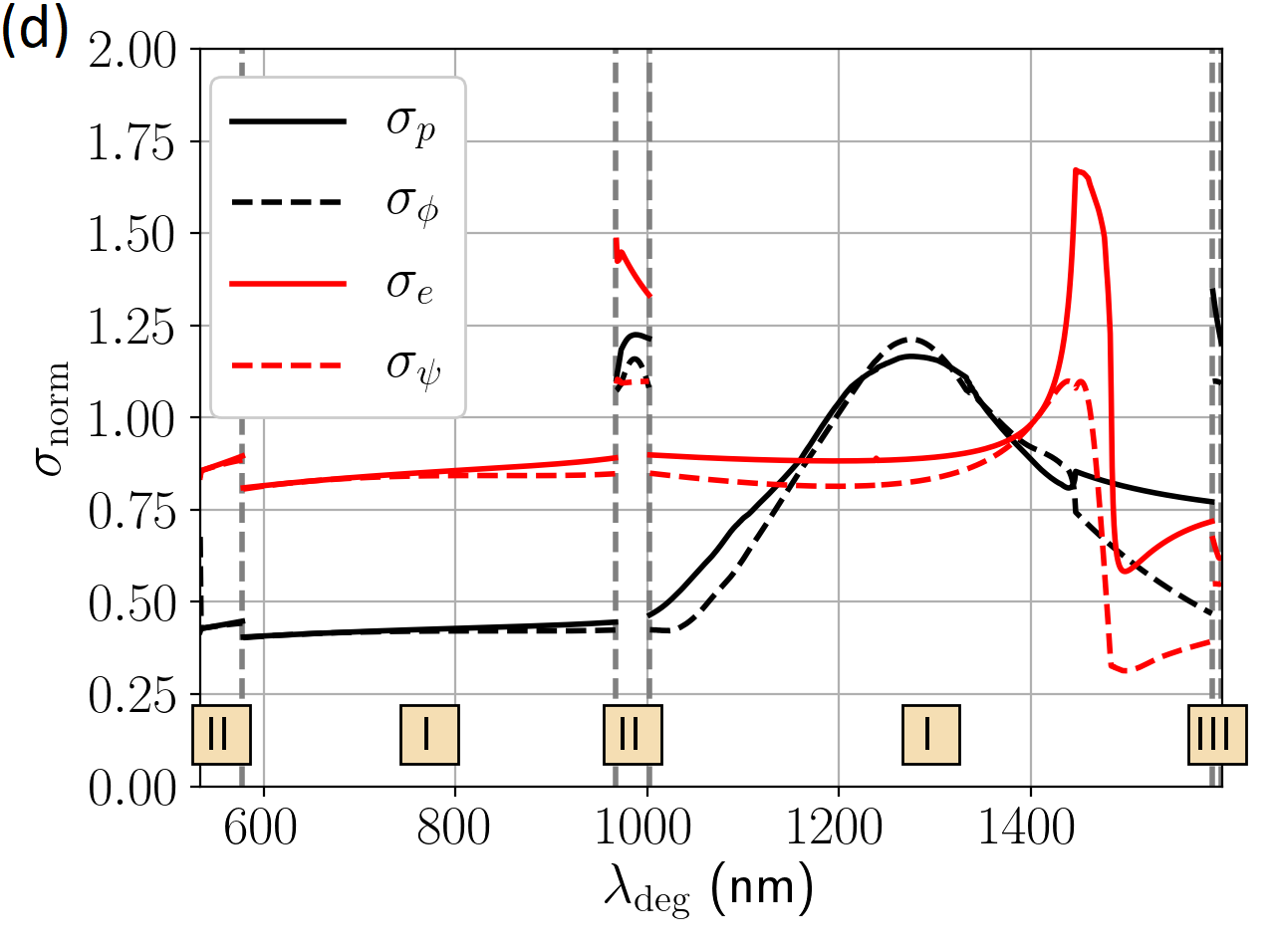}
	\label{fig sig norm vs lam LN}
	\end{subfigure}
	\begin{subfigure}{0.45\linewidth}	
	\includegraphics[width=\linewidth]{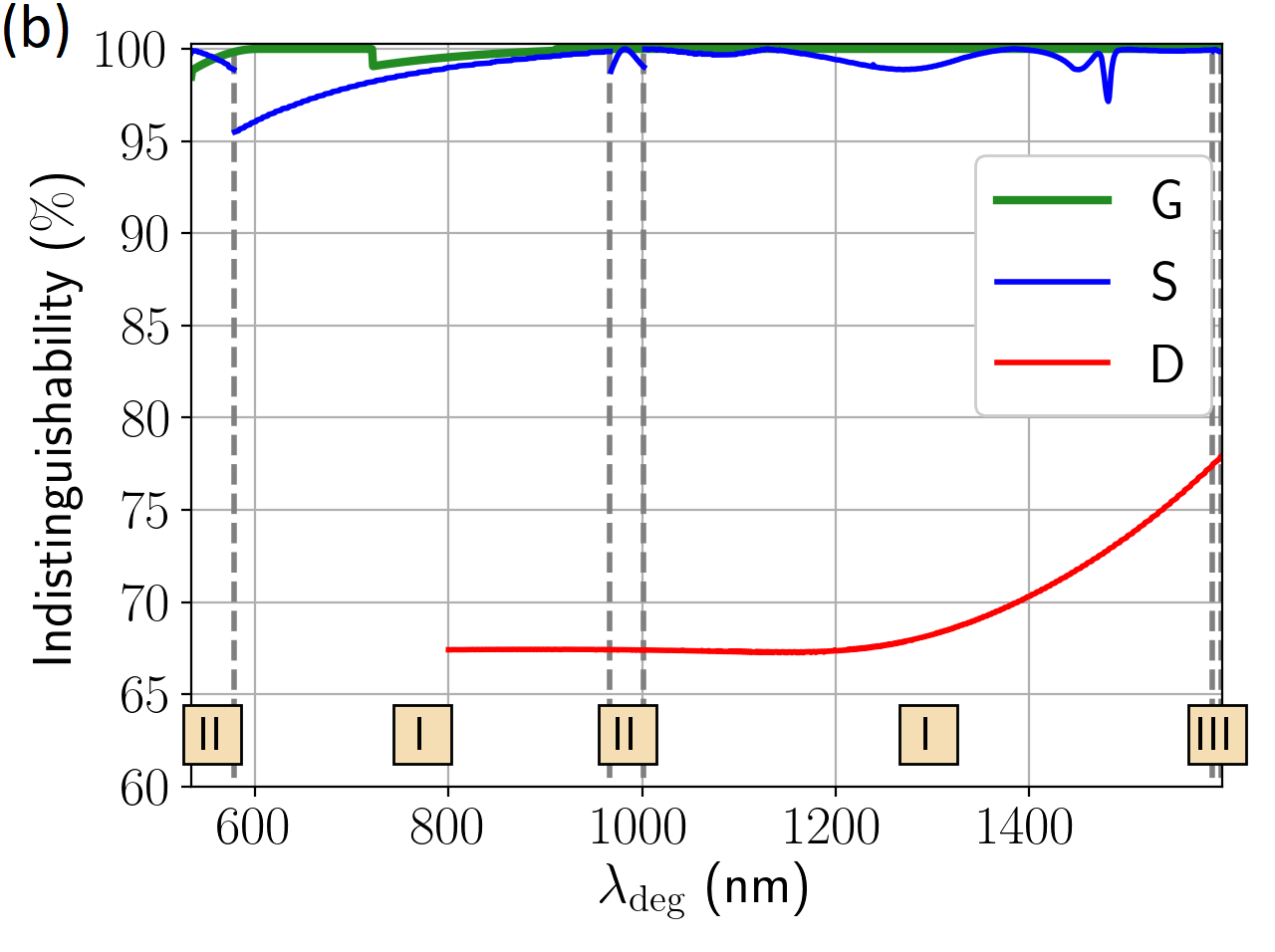}
	\label{fig I vs lam LN}
	\end{subfigure}%
	\begin{subfigure}{0.45\linewidth}	
	\includegraphics[width=\linewidth]{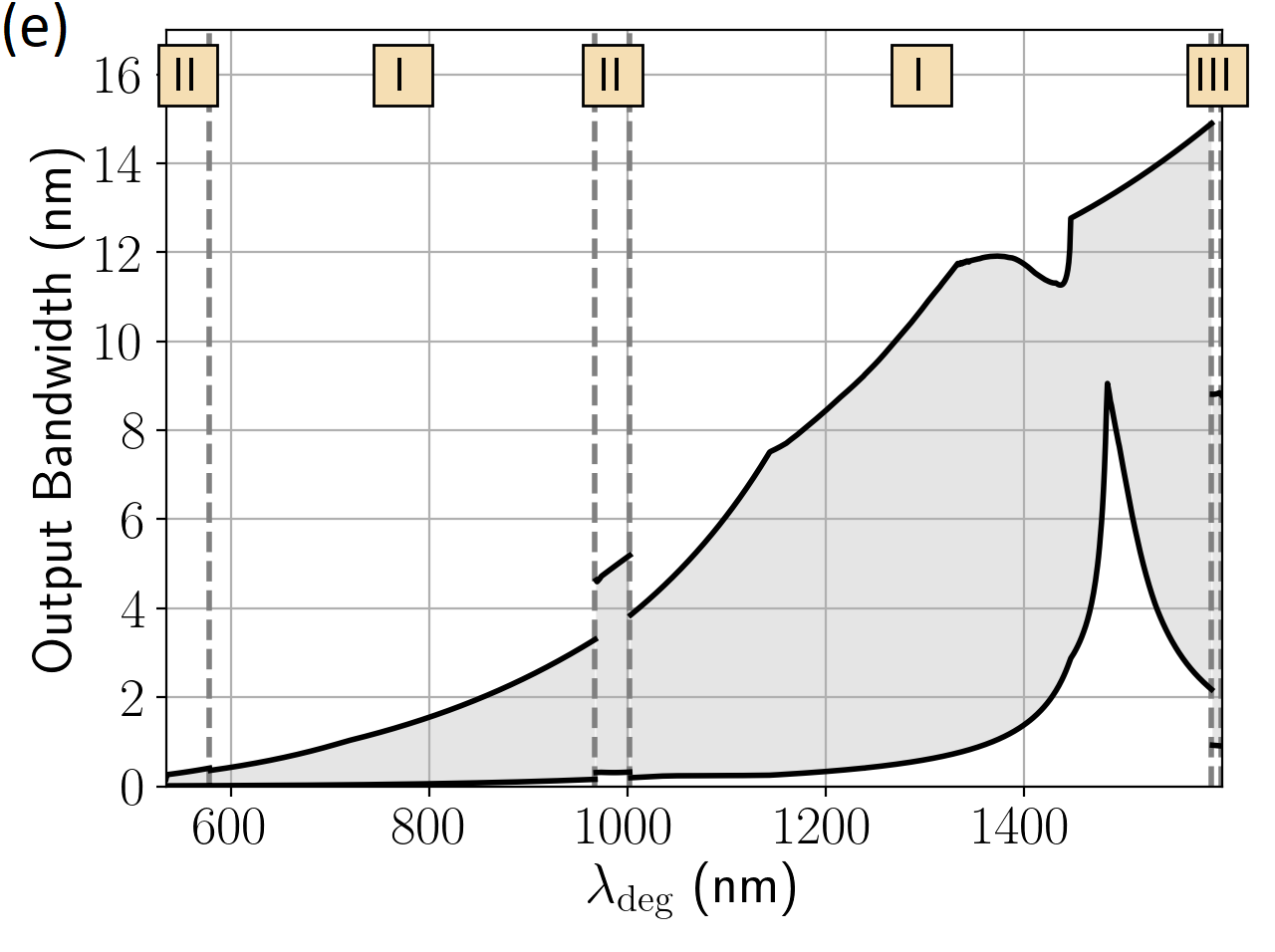}
	\label{fig sig out vs lam LN}
	\end{subfigure}
	\begin{subfigure}{0.45\linewidth}	
	\includegraphics[width=\linewidth]{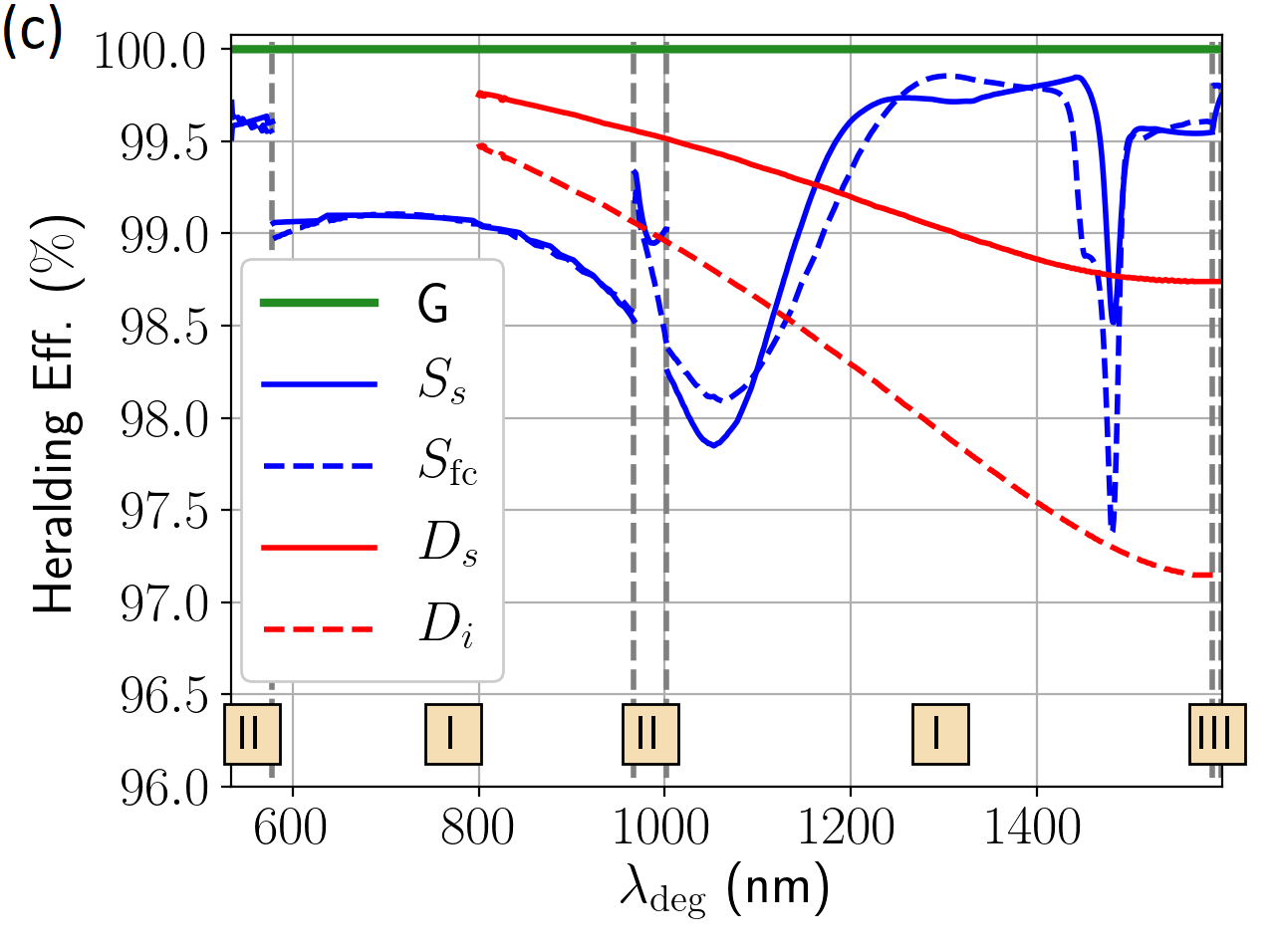}
	\label{fig H vs lam LN}
	\end{subfigure}%
	\begin{subfigure}{0.45\linewidth}	
	\includegraphics[width=\linewidth]{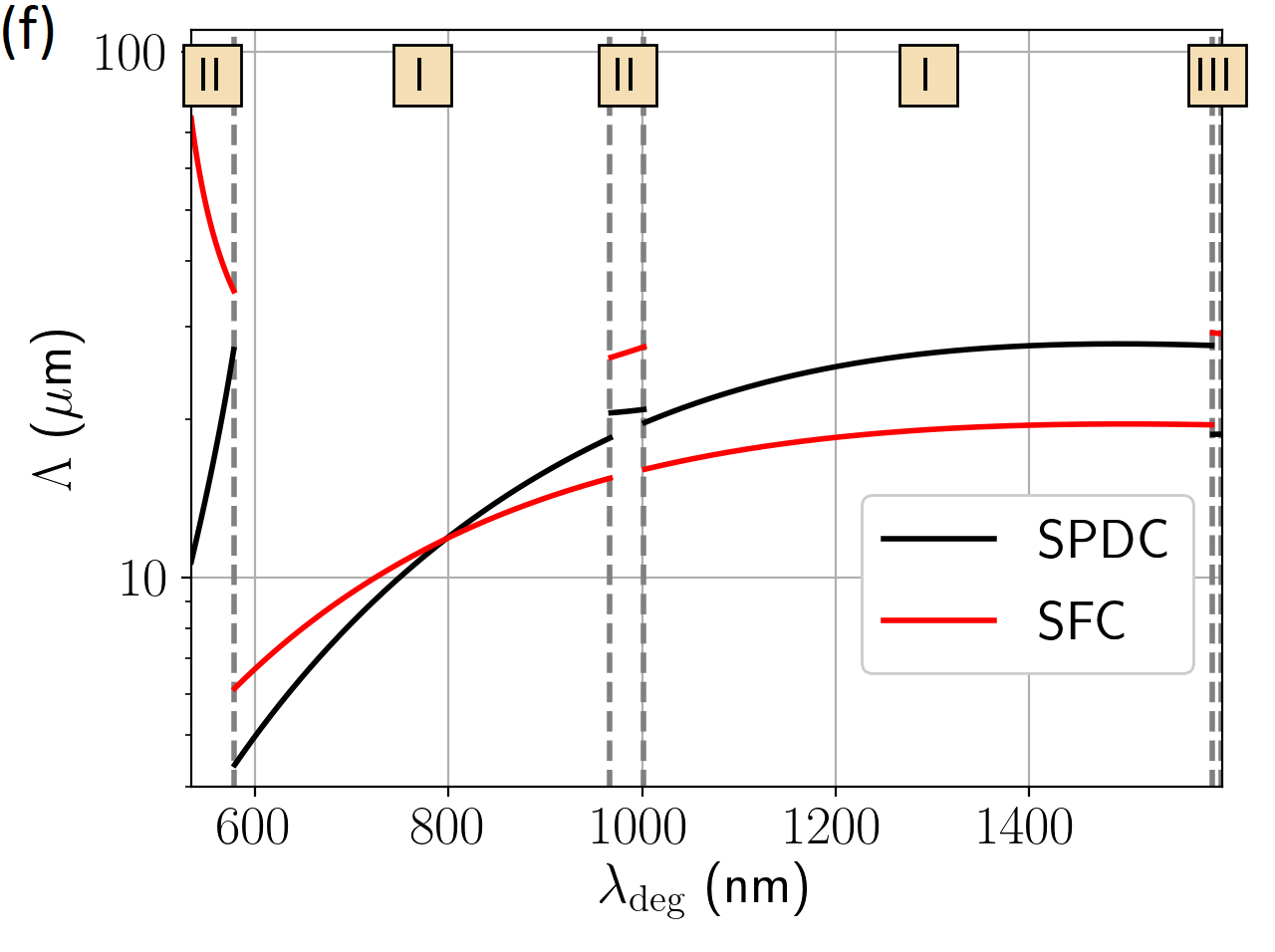}
	\label{fig Pol vs lam LN}
	\end{subfigure}%
	\caption{The results of the frequency-converted SPDC optimization for the nonlinear crystal LN.  The figure includes the (a) purity, the (b) indistinguishability, and (c) the heralding efficiency for frequency-converted SPDC with Gaussian G and sinc S phase matching, and for degenerate SPDC D with sinc phase matching (the curve S and D are filtered just enough to remove the sidebands from the JSA), the (d) optimal pump ($\sigma_p$), escort ($\sigma_e$), SPDC phase-matching bandwidth ($\sigma_\phi$), and SPC phase-matching bandwidth ($\sigma_\psi$) normalized to the target output SPDC bandwidth, the (e) achievable output SPDC bandwidths, and the (f) poling period for the SPDC and SFG regions of the crystal.  The optimal phase-matching configurations at each output wavelength ($\lambda_{\mathrm{deg}}$) are labeled by Roman numerals corresponding to the rows in Table \ref{tab: Config KTP}.  The boundaries are indicated by the vertical gray dashed lines. 
}
	\label{fig Result LN}
\end{figure*}
\begin{equation}
\label{equ: output band constraint}
1 \, \mathrm{mm} \leq L \leq 3 \, \mathrm{cm}
\end{equation}
on the individual lengths of the SPDC and SFC regions of the nonlinear crystal.
The lower and upper bounds of Eq. (\ref{equ: output band constraint}) define the upper and lower boundaries (black curves) in Fig.~\ref{fig Result KTP}(e), respectively.  
\begin{figure*}[t!]
	\begin{subfigure}{0.45\linewidth}
	\includegraphics[width=\linewidth]{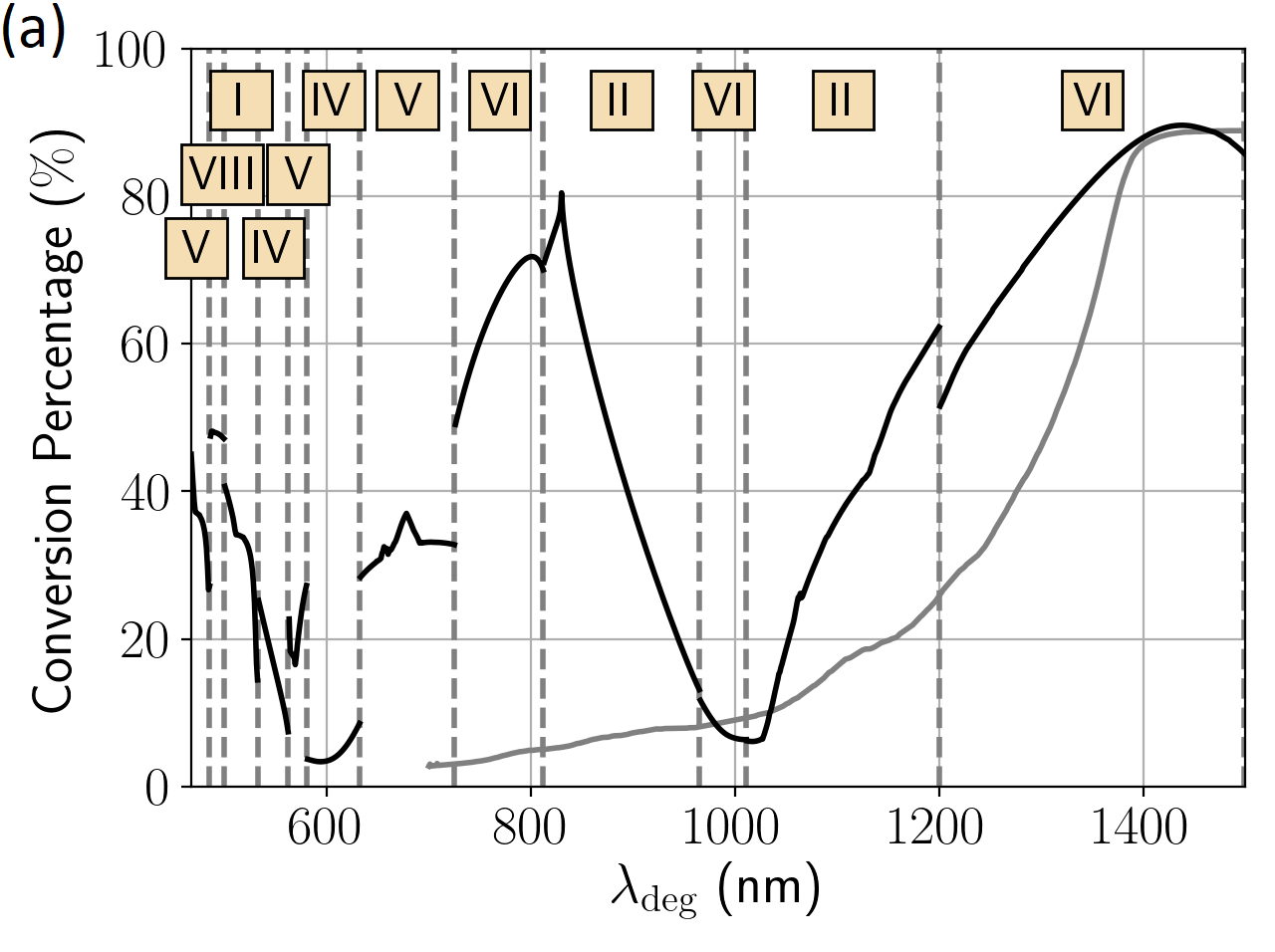}
	\label{fig Conv Perc KTP S}
	\end{subfigure}%
	\begin{subfigure}{0.45\linewidth}
	\includegraphics[width=\linewidth]{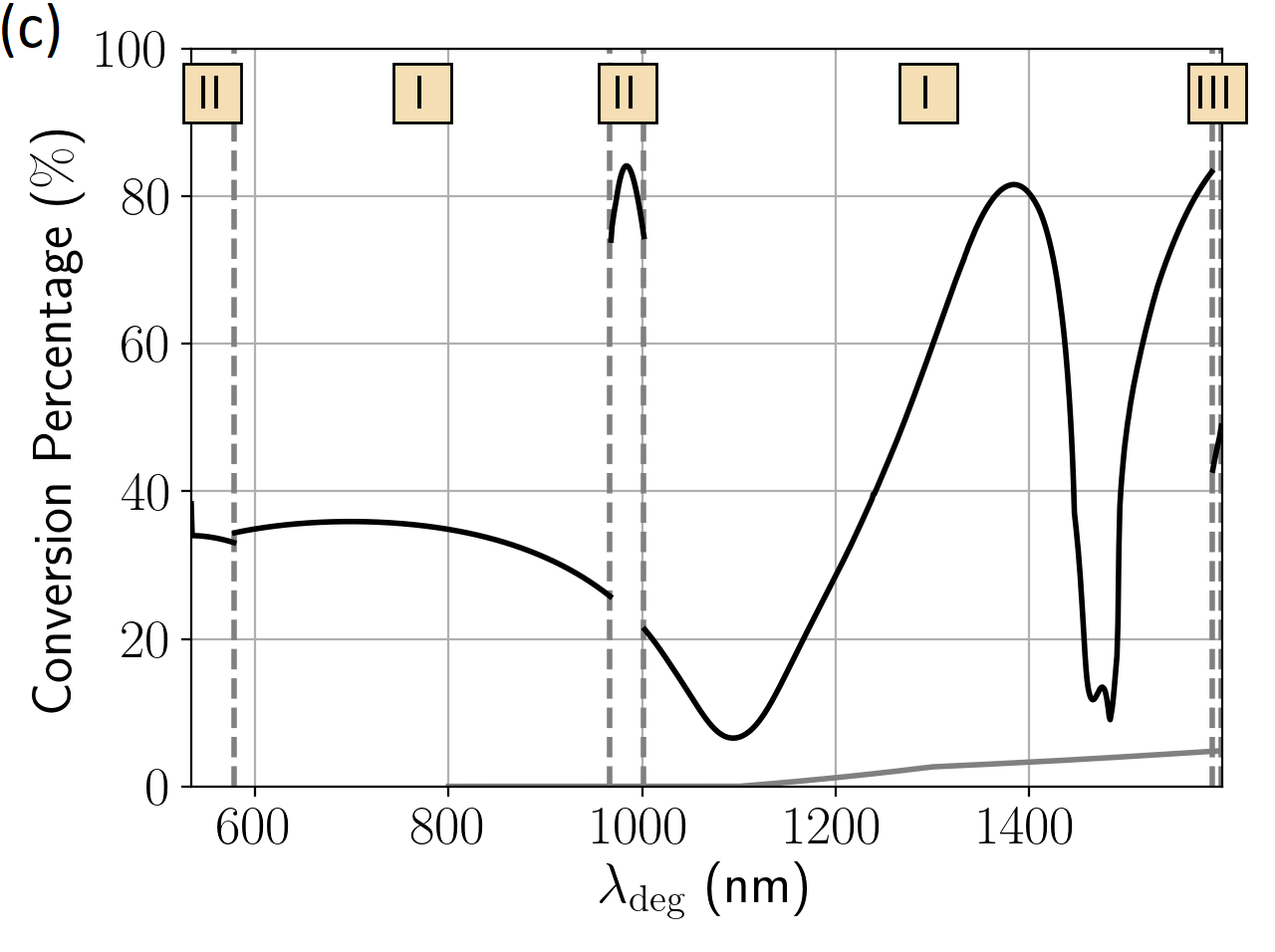}
	\label{fig Conv Perc LN S}
	\end{subfigure}
	\begin{subfigure}{0.45\linewidth}
	\includegraphics[width=\linewidth]{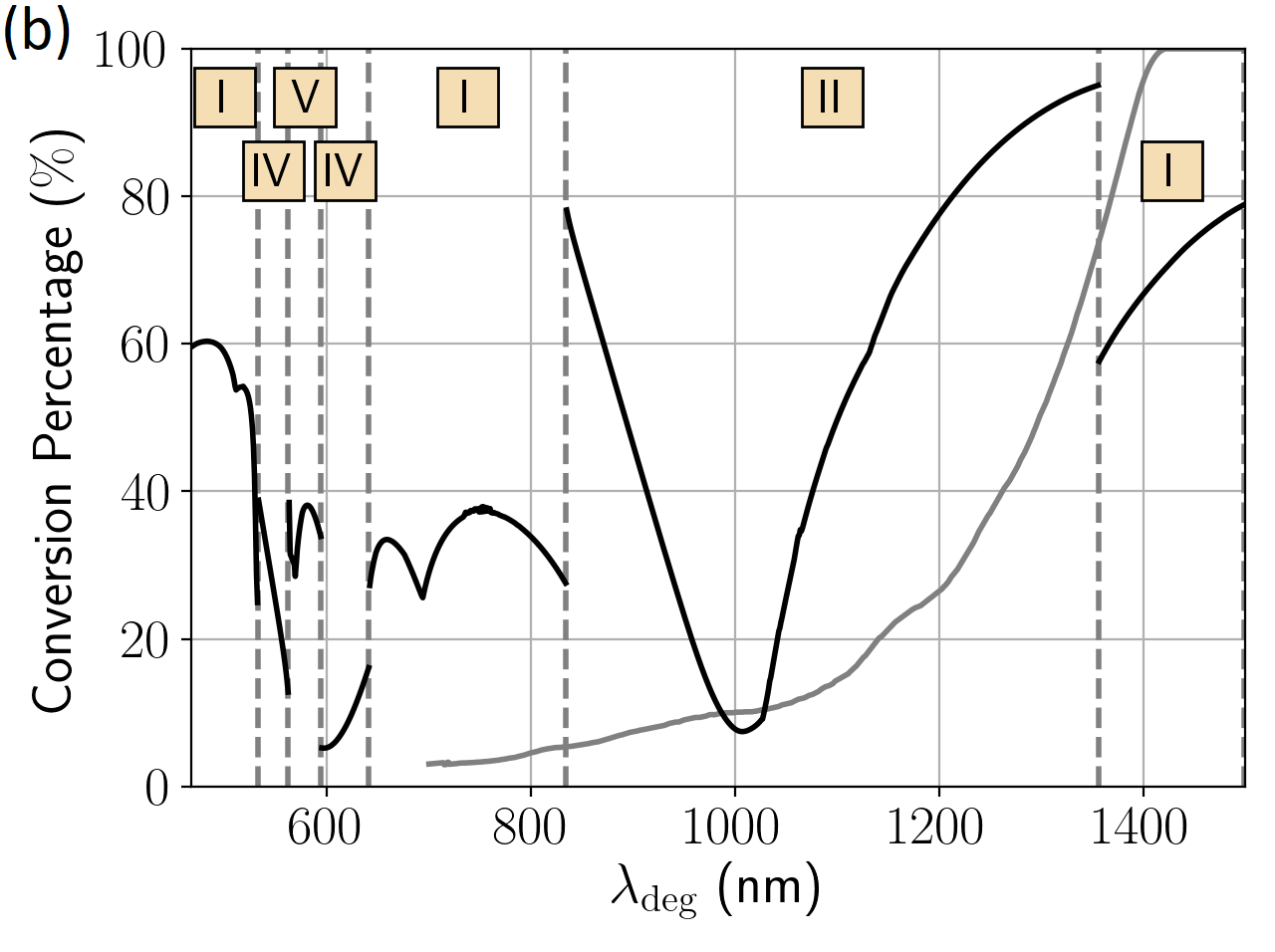}
	\label{fig Conv Perc KTP G}
	\end{subfigure}%
	\begin{subfigure}{0.45\linewidth}
	\includegraphics[width=\linewidth]{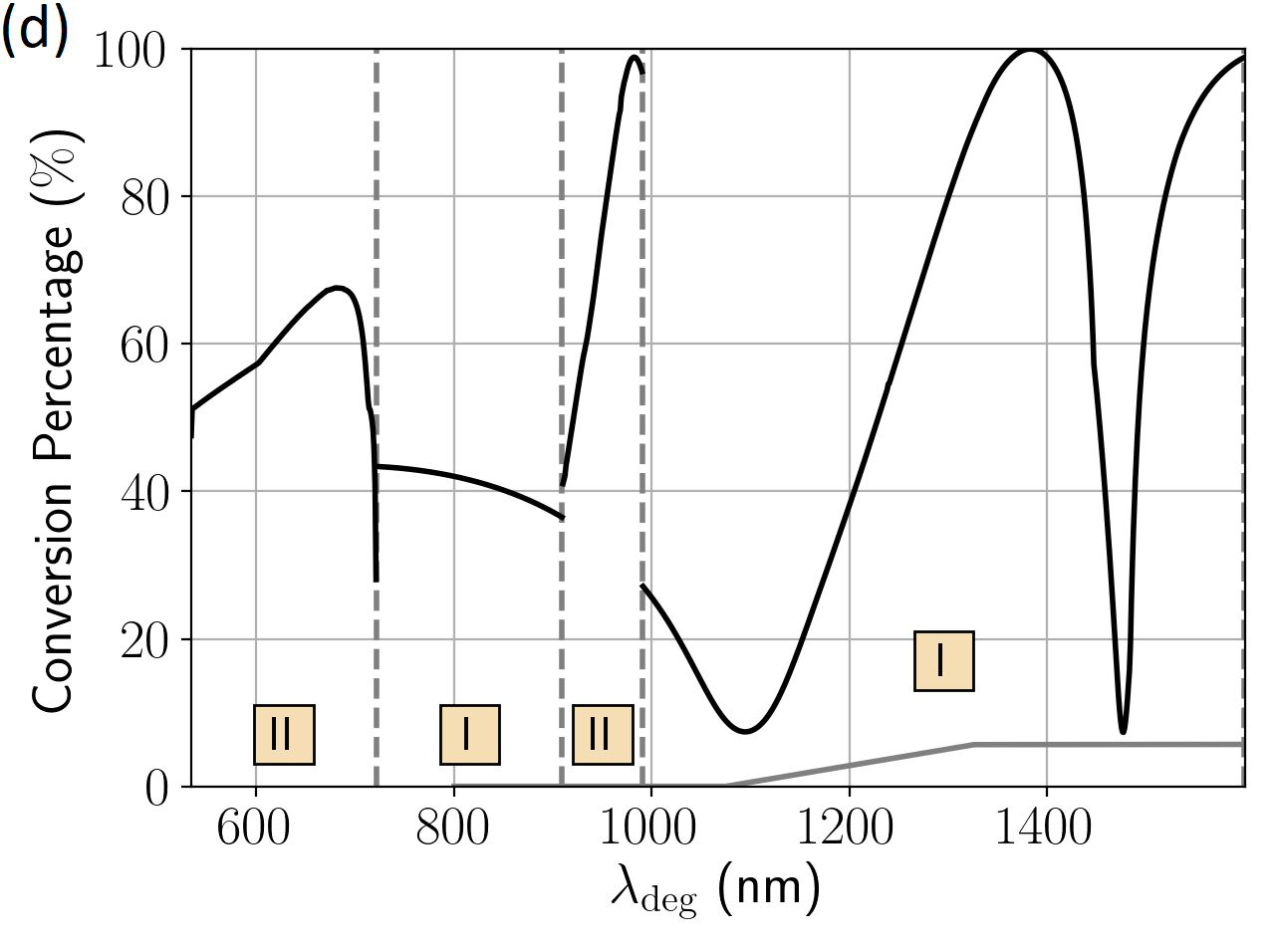}
	\label{fig Conv Perc LN G}
	\end{subfigure}%
	\caption{The conversion efficiency (black curves) for our FC-SPDC scheme using (a), (b)  KTP (see Fig.~\ref{fig Result KTP}) and (c), (d) LN (see Fig.~\ref{fig Result LN}) with (a), (c) sinc and (b), (d) Gaussian phase matching. For reference, the gray curves are the conventional filtering efficiency $\eta_{\mathrm{filt}} = | f_{\mathrm{filt}}| ^2/| f_{\mathrm{JSA}}| ^2$. The optimal phase-matching configuration (I--V) for each output wavelength ($\lambda_{\mathrm{deg}}$) is listed in Table \ref{tab: Config KTP} and the vertical dotted lines indicate their boundaries.
}
	\label{fig Conv Perc}
\end{figure*}
Figure~\ref{fig Result KTP}(a) shows that our FC-SPDC scheme achieves $P>96.5\%$ when using sinc phase matching (blue) for $466$ nm $\leq$ $\lambda_{\mathrm{deg}}$  $\leq$ $1500$ nm.
Except for the region $530$ nm $\leq$ $\lambda_{\mathrm{deg}}$  $\leq$ $560$ nm where $P>96.5\%$, the scheme achieves $P>99\%$ when using ideal Gaussian phase matching (green) [see Fig.~\ref{fig Purity Zoom}(a) in Appendix~\ref{sec:FC-SPDC Zoom} for more resolution of our FC-SPDC results].  
For comparison, the degenerate SPDC process has $P$ approaching unity (after filtering away the sinc sidebands) as $\lambda_{\mathrm{deg}}\rightarrow1550$ nm, but it quickly diverges from unity as $\lambda_{\mathrm{deg}}$ decreases.  
The degenerate SPDC curve terminates below 710 nm because KTP begins to absorb at the pump wavelength (this is discussed further at the end of this section).  
Figure~\ref{fig Result KTP}(b) shows $I$ is close to $100\%$ except for several regions where the phase-matching angles become unfavorable and $I$ drops as low as $91\%$. 
Figure~\ref{fig Result KTP}(c) gives the heralding efficiency $H$ for the case of sinc phase matching and filtering just enough to suppress the sidebands without truncating the central lobe of $f_{\mathrm{eff}}$.  
Our scheme allows one to independently increase the escort power in an effort to approach near unit conversion efficiency of the overlapping spectral regions \cite{brecht2011quantum}, and thus conversion efficiency has been neglected in the present analysis. 
A detailed analysis of the effect of nonunit conversion efficiency on the heralding efficiency and other quantities is reserved for future work.
Considering the metrics together, we see for $466$ nm $\leq$ $\lambda_{\mathrm{deg}}$  $\leq$ $1500$ nm one can simultaneously achieve ultrahigh purity $P$ and indistinguishability $I$, while maintaining high heralding efficiency $H$ and using broad spectral filters.

Figure~\ref{fig Result KTP}(d) shows the bandwidth parameter optimization, where $\sigma_p$, $\sigma_\phi$, $\sigma_e$, and $\sigma_\psi$ are the normalized bandwidths of $\alpha$, $\Phi$, $\beta$, and $\Psi$, respectively, which are normalized to the target output bandwidth of the source.  
We force the following constrains on the bandwidth optimization algorithm to ensure the algorithm searches over a realistic parameter space: $\sigma_p\leq2\sigma_e$, $\sigma_e\leq2\sigma_p$, $\sigma_\phi\leq2\sigma_\psi$, $\sigma_\psi\leq2\sigma_\phi$, and $\sigma_p$, $\sigma_\phi$, $\sigma_e$, and $\sigma_\psi$ to be less than 6 and greater than 0.1.
In addition, we require the temporal duration $\delta t$ of the pump and escort beams to be 5 fs $\leq$ $\delta t$ $\leq$ 1 ns.
Figure~\ref{fig Result KTP}(e) shows the available target output bandwidths (shaded gray region) where we assert these constraints and Eq. (\ref{equ: output band constraint}). 
Figure~\ref{fig Result KTP}(f) shows the required poling period for quasi-phase-matching of the SPDC process (black) and the SFC process (red) with the phase-matching configuration indicated by I through VIII (see Table \ref{tab: Config KTP}). 
The peak at $\lambda_{\mathrm{deg}}\approx 514$ nm in the black curve indicates the nondegenerate SPDC process is almost phase matched at this $\lambda_{\mathrm{deg}}$.

Similarly, Fig.~\ref{fig Result LN} shows the result of the optimization when using the nonlinear crystal LN. 
Figures~\ref{fig Result LN}(a)--\ref{fig Result LN}(c) indicate our FC-SPDC scheme can achieve $P>95.5\%$, $I>95\%$, and $H>97\%$ for $534$~nm $\leq$ $\lambda_{\mathrm{deg}}$  $\leq$ $1600$~nm when using sinc phase matching  [see Fig.~\ref{fig Purity Zoom}(b) in Appendix \ref{sec:FC-SPDC Zoom} for more resolution].  
If one can achieve perfect Gaussian phase matching, $P$ increases to $100\%$, $I$ increases to greater than $98\%$, and $H=100\%$ because there are no sidebands and therefore no required filtering.  
The simulations are similar when using the nonlinear crystal magnesium oxide doped lithium niobate (MgLN), and the reader is referred to the LN simulations if wanting to use MgLN.

Another benefit of our FC-SPDC scheme is the ability to create polarization entangled photons at $\lambda_{\mathrm{deg}}$ below what is possible when using conventional degenerate SPDC processes.
This is because the KTP and LN or MgLN absorb below $\lambda_{\mathrm{deg}}\approx355$ nm \cite{KTPtrans} and $\lambda_{\mathrm{deg}}\approx400$ nm \cite{LNtrans}, respectively.
Degenerate SPDC obeys the relationship $\lambda_{\mathrm{deg}}=2\lambda_p$, and thus when using KTP and LN the lower limit for SPDC is $\lambda_{\mathrm{deg}}\approx710$ nm and $\lambda_{\mathrm{deg}}\approx800$ nm, respectively.
Our FC-SPDC source obeys $\lambda_{\mathrm{deg}}=4\lambda_p/3$, and therefore the lower limit is $\lambda_{\mathrm{deg}}\approx466$ nm for KTP and $\lambda_{\mathrm{deg}}\approx534$ nm for LN.
Consequently, if one desires to operate in the short-wavelength region of the visible spectrum that, for example, is optimal for free-space quantum communication \cite{lanning2021quantum}, our FC-SPDC scheme is a viable option whereas conventional-degenerate SPDC is not.

Since our scheme is analogous to a filtering strategy, the conversion efficiency is analogous to the probability that both photons are transmitted through a spectral filter (see $\mathcal{P}_{\mathrm{both}}^{(G)}$ and $\mathcal{P}_{\mathrm{both}}^{(s)}$ in Fig.~\ref{fig Deg KTP Filter}).
In fact, noting Eqs. (\ref{eq:P_BOTH}) and (\ref{eq:conv_eff}), one will see they are mathematically identical.
In Fig.~\ref{fig Conv Perc} we plot the conversion efficiency $\eta_{\mathrm{conv}}$ given in Eq. (\ref{eq:conv_eff}) (black curve) and the conventional filtering efficiency $\mathcal{P}_{\mathrm{both}}^{(i)}$ given in Eq. (\ref{eq:P_BOTH}) (gray curve)).
For KTP [Figs. \ref{fig Conv Perc}(a) and \ref{fig Conv Perc}(c)], one can see that $\eta_{\mathrm{conv}}$ is considerably larger than $\mathcal{P}_{\mathrm{both}}^{(i)}$ for most of the wavelength range except for a narrow region near 1000 nm and near 1400 nm where conventional phase matching becomes intrinsically pure.
For LN [Figs. \ref{fig Conv Perc}(b) and \ref{fig Conv Perc}(d)], one can see $\eta_{\mathrm{conv}}$ is considerably larger for most regions and in this case the conventional filtering efficiency $\mathcal{P}_{\mathrm{both}}^{(i)}$ plateaus near 5$\%$.
Other inefficiencies related to spatial and transverse-spatial mode overlap are inherent to both methods, but are neglected here.

\section{Conclusion}
\label{sec: Conclusion}
In this paper we present a technique to generate ultrahigh purity, indistinguishable, degenerate, polarization entangled photon pairs over a large visible and NIR spectral range.  
The technique uses a single nonlinear crystal segmented into three phase-matching regions.  
The central region is quasi-phase-matched for nondegenerate SPDC and the leading and trailing regions are identically quasi-phase-matched for SFC.  
The idler photon from an SPDC event is frequency converted into the same frequency mode as the signal photon by means of a high-energy pulsed escort beam.  
Therefore, we call this two-step process frequency-converted SPDC (FC-SPDC).  
Furthermore, the crystal is placed at the center of a Sagnac interferometer to generate frequency-uncorrelated polarization entanglement between the output modes of the interferometer.  

We show, by carefully selecting the lengths of the SPDC and SFC regions, the phase-matching configuration (listed in Table \ref{tab: Config KTP}), and the nonlinear crystal KTP, or LN, it is possible to achieve heralded-single-photon spectral purity exceeding $99\%$, indistingishability above $95\%$, and heralding efficiency above $94\%$ for all degenerate operating wavelengths between approximately $466$ nm and $1600$ nm.  
This scheme is also a good source of high-quality polarization-entangled photon pairs below approximately equal to $710$ nm, which is difficult to achieve by means of conventional-degenerate SPDC due to absorption of the pump.  
Considering these metrics, our FC-SPDC scheme is an example of an ideal source for a polarization-based quantum network.  
However, further work is required to model the brightness of FC-SPDC and the transverse-spatial-mode structure. 


\section{Acknowledgments}
R.L. and R.N.L. acknowledge support from the National Research Council (NRC) Research Associateship Programs (RAP) and the Office of the Secretary of Defense (OSD) ARAP Defense Optical Channel Program (DOC-P), respectively.
The authors gratefully acknowledge support from the DOC-P principal investigator Mark T. Gruneisen, the program manager Valerie Knight, and helpful technical discussions with AdvR Inc.

The views expressed are those of the authors and do not reflect the official guidance or position of the U.S. Government, the Department of Defense, or of the U.S. Air Force.

The appearance of external hyperlinks does not constitute endorsement by the U.S. Department of Defense (DoD) of the linked websites, or the information, products, or services contained therein. The DoD does not exercise any editorial, security, or other control over the information you may find at these locations.

Approved for public release; distribution is unlimited. Public Affairs release approval AFRL-2021-4144.

\appendix

\section{}\label{sec:FC-SPDC Zoom}
Here we collect complementary figures that are too numerous for the main text, which includes analysis of KTP and LN.
In Appendix \ref{sec:OptForGaussianKTPandLN} we present the results of the optimization procedure when using Gaussian phase matching for both KTP and and LN.
In Appendix \ref{sec:FC-SPDC Zoom} we decrease the scale of the purity plots in order to give more resolution of the performance of our scheme.

\begin{figure}[!]
	\begin{subfigure}{0.93\linewidth}
	\includegraphics[width=\linewidth]{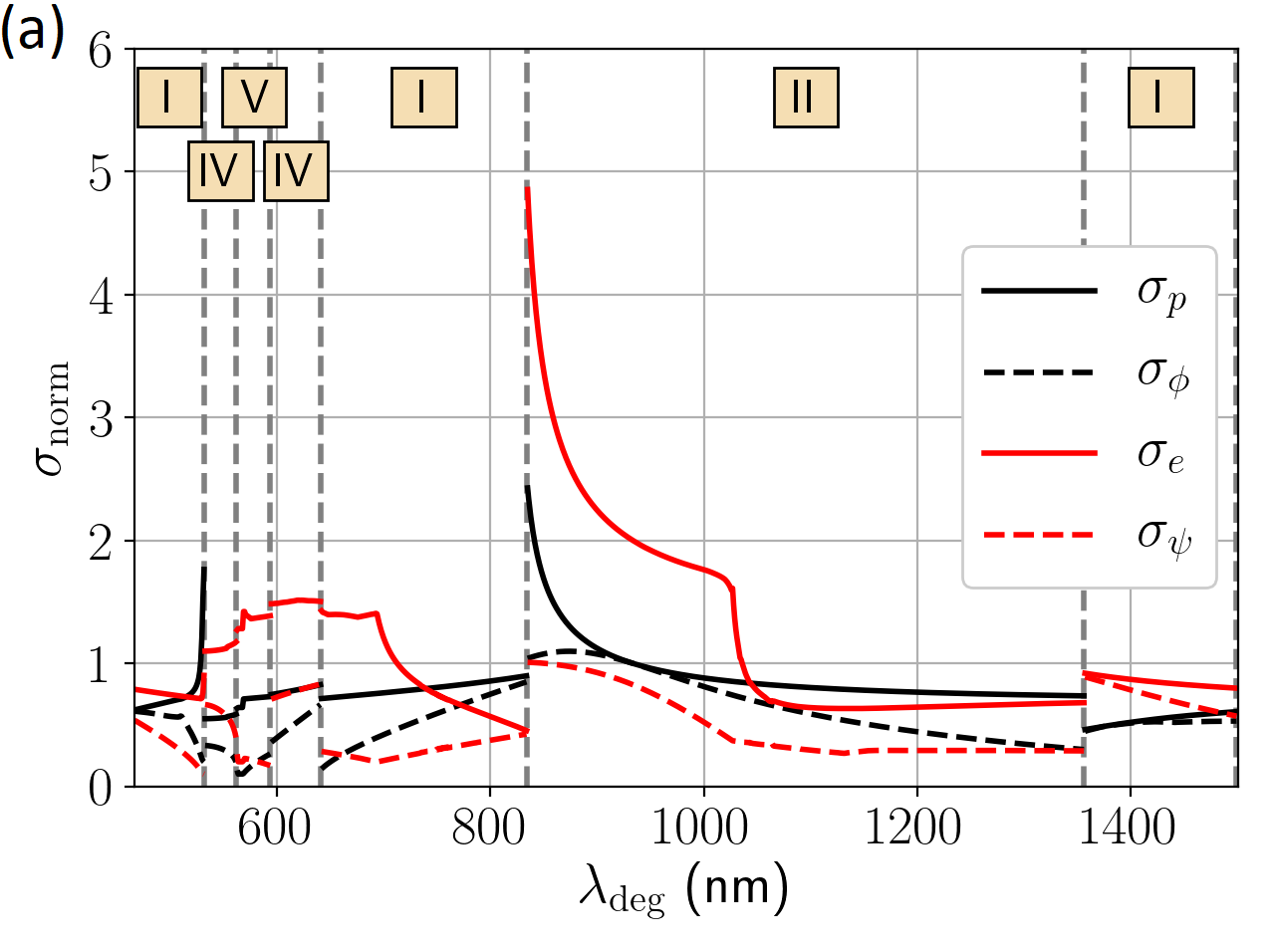}
	\label{fig sig norm vs lam KTP gauss}
	\end{subfigure}
	\begin{subfigure}{0.93\linewidth}
	\includegraphics[width=\linewidth]{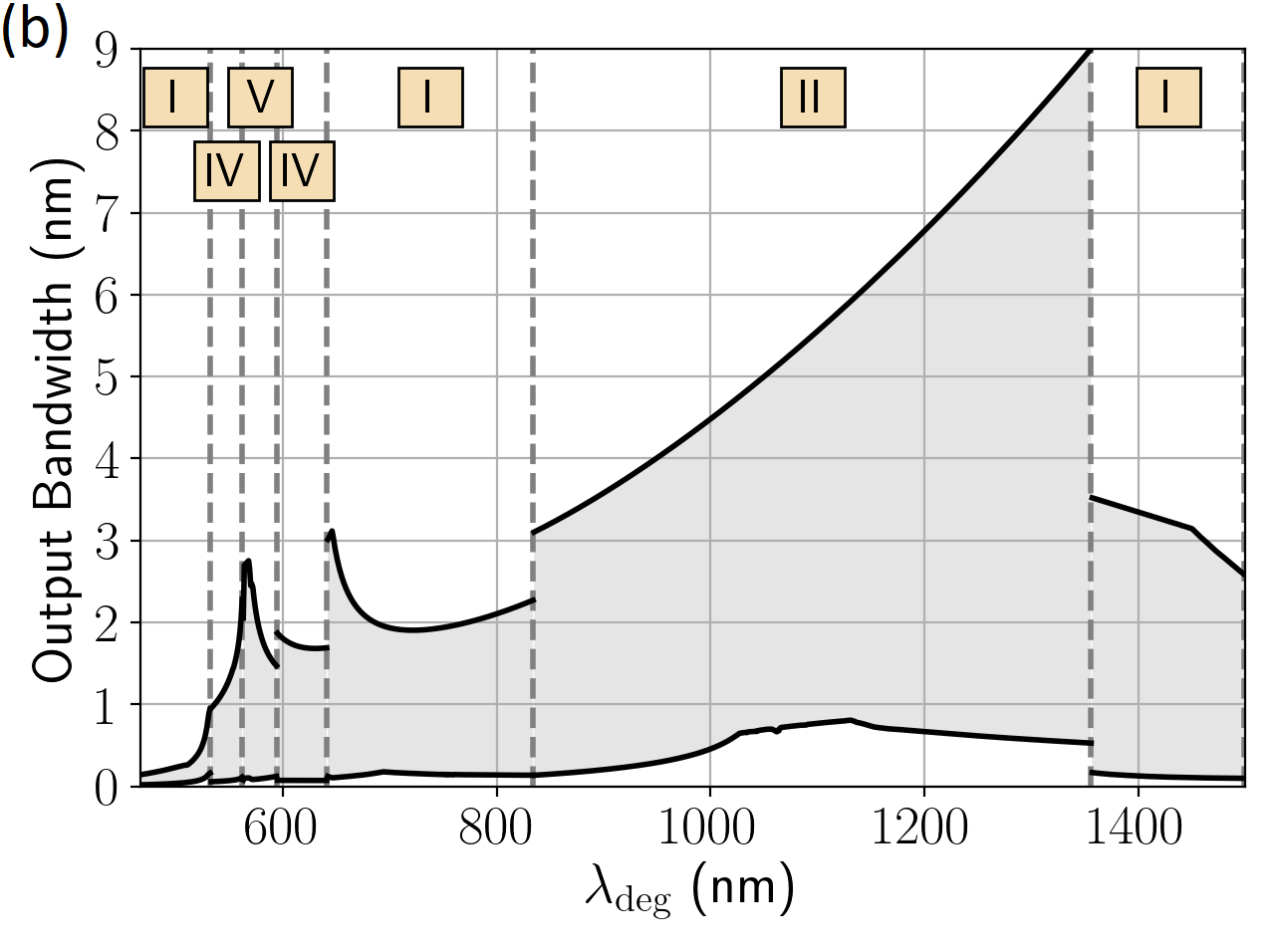}
	\label{fig sig out vs lam KTP gaus}
	\end{subfigure}
	\begin{subfigure}{0.93\linewidth}
	\includegraphics[width=\linewidth]{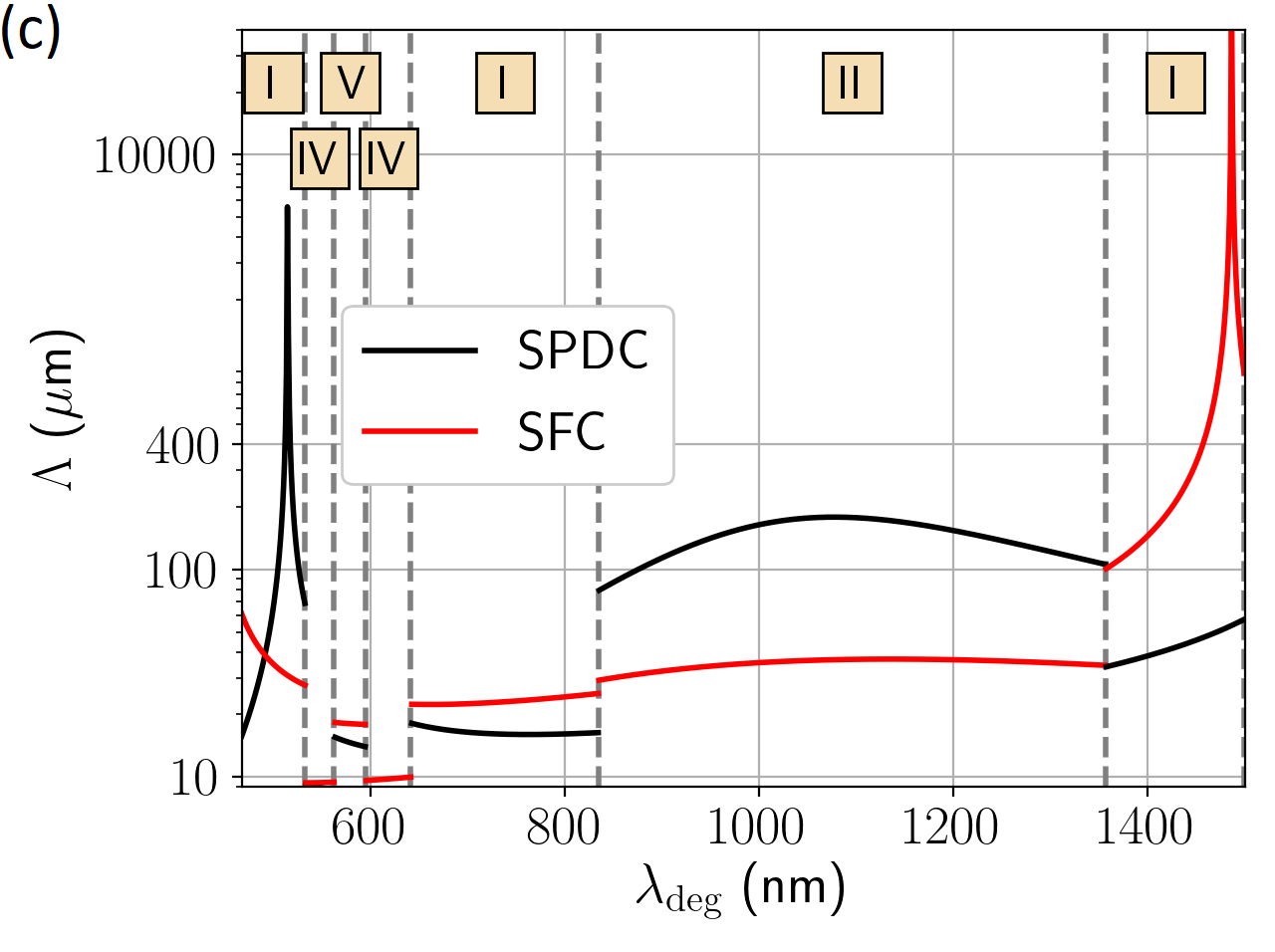}
	\label{fig Pol vs lam KTP gaus}
	\end{subfigure}	
	\caption{Results of the optimization when using the nonlinear crystal KTP and Gaussian phase matching including (a) the optimal pump ($\sigma_p$), escort ($\sigma_e$), SPDC phase-matching bandwidth ($\sigma_\phi$), and SPC phase-matching bandwidth ($\sigma_\psi$) normalized to the target output SPDC bandwidth, the (b) achievable output SPDC bandwidths, and the (c) poling period for the SPDC and SFG regions of the crystal.  The optimal phase-matching configurations at each output wavelength ($\lambda_{\mathrm{deg}}$) are labeled by Roman numerals corresponding to the rows in Table \ref{tab: Config KTP}.  The boundaries are indicated by the vertical gray dashed lines.
}

	\label{fig Gaus Results KTP}
\end{figure}

\begin{figure}[!]
	\begin{subfigure}{0.93\linewidth}
	\includegraphics[width=\linewidth]{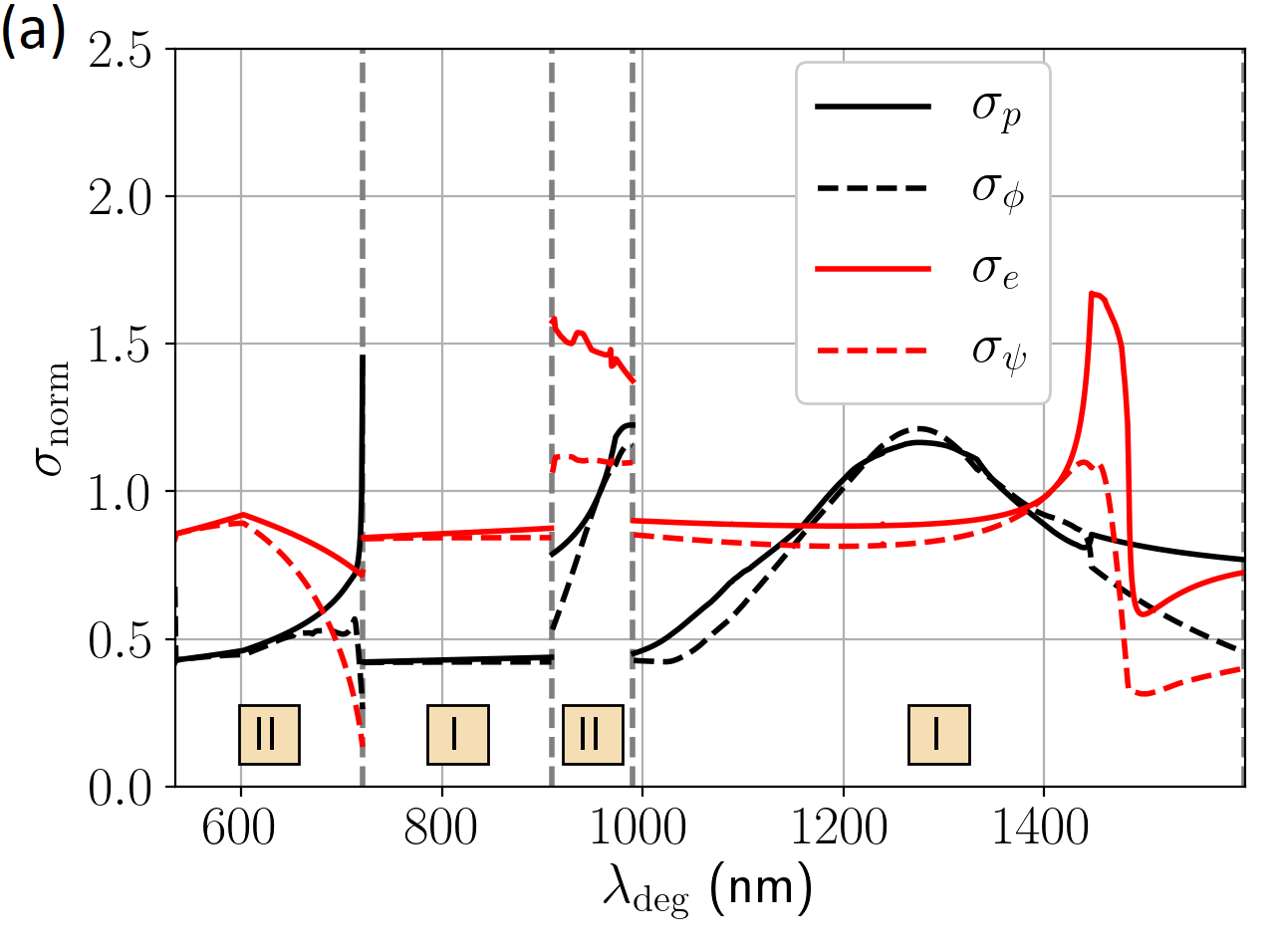}
	\label{fig sig norm vs lam LN gauss}
	\end{subfigure}
	\begin{subfigure}{0.93\linewidth}
	\includegraphics[width=\linewidth]{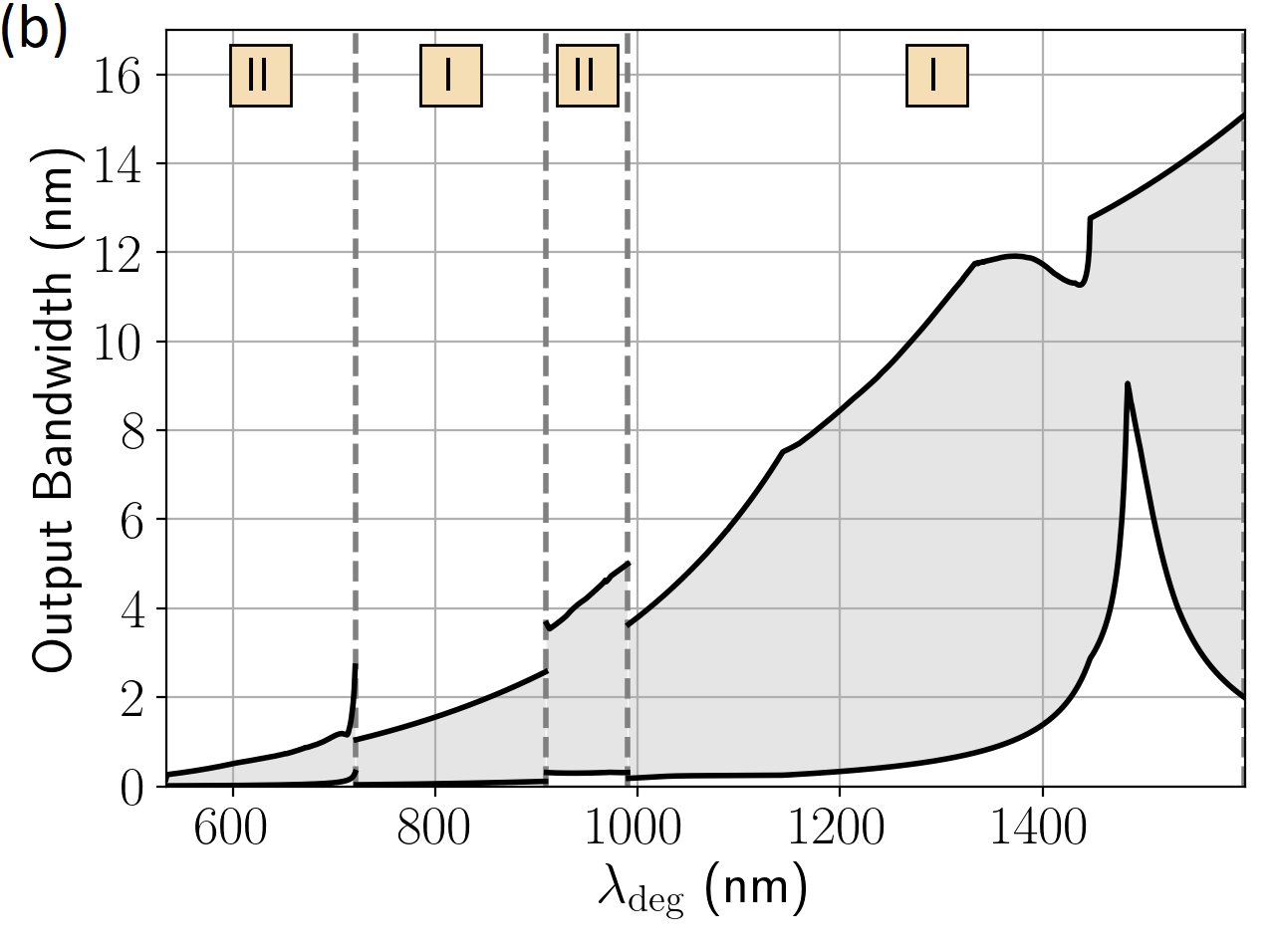}
	\label{fig sig out vs lam LN gaus}
	\end{subfigure}
	\begin{subfigure}{0.93\linewidth}
	\includegraphics[width=\linewidth]{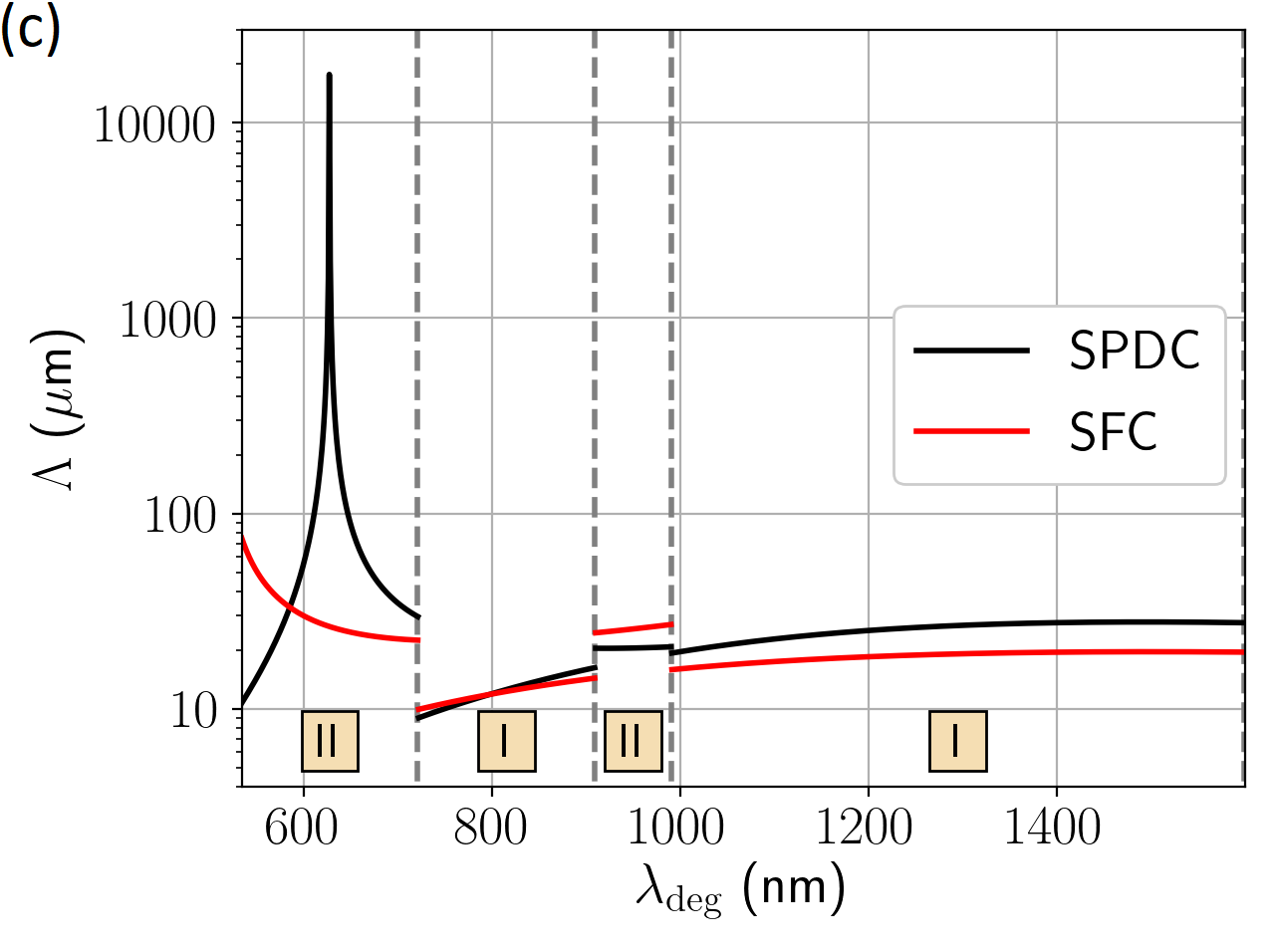}
	\label{fig Pol vs lam LN gaus}
	\end{subfigure}
	\caption{Results of the optimization when using the nonlinear crystal LN and Gaussian phase matching including (a) the optimal pump ($\sigma_p$), escort ($\sigma_e$), SPDC phase-matching bandwidth ($\sigma_\phi$), and SPC phase-matching bandwidth ($\sigma_\psi$) normalized to the target output SPDC bandwidth, the (b) achievable output SPDC bandwidths, and the (c) poling period for the SPDC and SFG regions of the crystal.  The optimal phase-matching configurations at each output wavelength ($\lambda_{\mathrm{deg}}$) are labeled by Roman numerals corresponding to the rows in Table \ref{tab: Config KTP}.  The boundaries are indicated by the vertical gray dashed lines.
}
	\label{fig Gaus Results LN}
\end{figure}

\begin{figure*}[!]
	\begin{subfigure}{0.49\linewidth}
	\includegraphics[width=\linewidth]{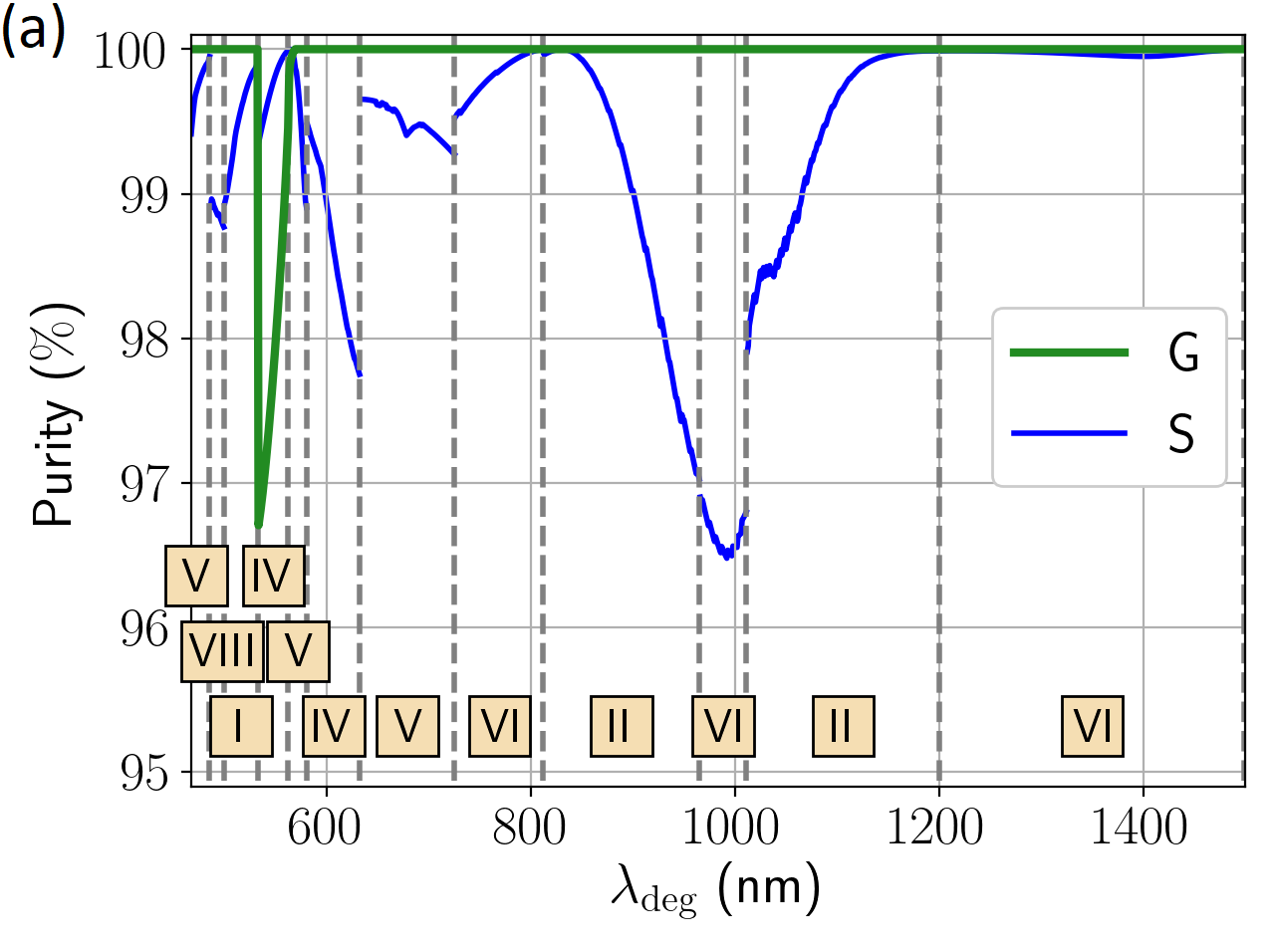}
	\label{fig P KTP closeup}
	\end{subfigure}%
	\begin{subfigure}{0.49\linewidth}
	\includegraphics[width=\linewidth]{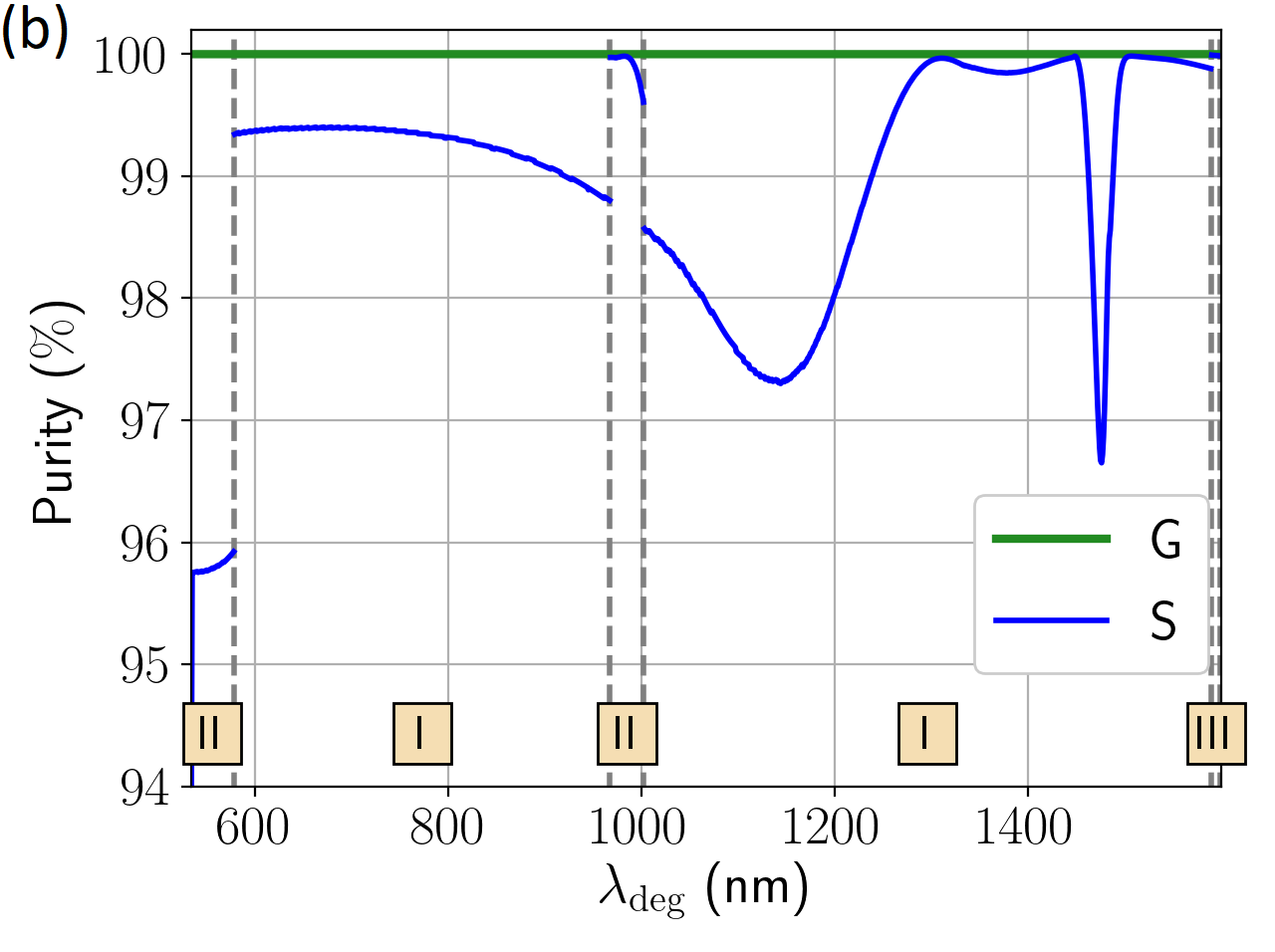}
	\label{fig P LN closeup}
	\end{subfigure}
	\caption{Reduced-range view of the purity of our FC-SPDC scheme for (a) KTP and (b) LN, corresponding to Figs. \ref{fig Result KTP}(a), and \ref{fig Result LN}(a), respectively.
}
	\label{fig Purity Zoom}
\end{figure*} 

\subsection{Optimization for Gaussian PMF}\label{sec:OptForGaussianKTPandLN}
In the main text we present the optimization parameters for the case of phase matching without domain engineering, which leads to a sinc PMF.
Due to the added complication of domain engineering, we reserve the optimization parameters for Gaussian phase matching to the Appendix and discuss them now.
Figures \ref{fig Gaus Results KTP}(a)--\ref{fig Gaus Results KTP}(c) show the  bandwidth parameters ($\sigma_p$, $\sigma_\phi$, $\sigma_e$, and $\sigma_\psi$), the range of available output bandwidths of the source, and the poling period, respectively, when using KTP.
Similarly, Figs. \ref{fig Gaus Results LN}(a)--\ref{fig Gaus Results LN}(c) show the parameters when using LN.
The metrics P, I, and H that result from these optimal parameters are plotted as the green curves in Figs. \ref{fig Result KTP}(a)--\ref{fig Result KTP}(c) and Figs \ref{fig Result LN}(a)--\ref{fig Result KTP}(c) when using KTP and LN, respectively. 

\subsection{Reduced-range view}\label{sec:FC-SPDC Zoom}
The physics of conventional-degenerate phase matching required a large range of values on the $y$ axis because the purity and indistinguishability can be so low. 
In Fig.~\ref{fig Purity Zoom} we reduce the range to show more resolution of our results from Figs.~\ref{fig Result KTP}(a) and \ref{fig Result LN}(a).
The green curves (G) and blue curves (S) indicate Gaussian and sinc phase matching, and the Roman numerals I through VIII indicate the optimal phase-matching configuration from Table \ref{tab: Config KTP}.
The vertical, gray lines mark the boundary of each phase-matching configuration.


\bibliographystyle{apsrev}
\bibliography{bibliography/bib_manuscript_eng_tunable_type2_deg_pol_ent_high_purity_source_v2}

\end{document}